\documentclass[12pt,epsbox]{article}
\usepackage[dvips]{graphicx}
\usepackage{amsmath}

\baselineskip=\normalbaselineskip

\newcommand{\resection}[1]{\setcounter{equation}{0}\section{#1}}
\renewcommand{\theequation}{\thesection.\arabic{equation}}
\renewcommand{\thefootnote}{\fnsymbol{footnote}}
\newcommand{\bel}[1]{\begin{equation}\label{#1}}
\newcommand{\bal}[1]{\begin{eqnarray}\label{#1}}

\newcommand{\be}{\begin{equation}}
\newcommand{\ee}{\end{equation}}
\newcommand{\ba}{\begin{eqnarray}}
\newcommand{\ea}{\end{eqnarray}}
\newcommand{\nn}{\nonumber \\}

\newcommand{\NP}[3]{Nucl. Phys. {\bf #1} {(#2)} {#3}}
\newcommand{\PL}[3]{Phys. Lett. {\bf #1} {(#2)} {#3}}
\newcommand{\IJMP}[3]{Int. J. Mod. Phys. {\bf #1} {(#2)} {#3}}
\newcommand{\PR}[3]{Phys. Rev. {\bf #1} {(#2)} {#3}}
\newcommand{\n}{\nonumber}

\newcommand{\bR}{{\bf R}}
\newcommand{\bZ}{{\bf Z}}
\newcommand{\hg}{\widehat{g}}

\newcommand{\eq}[1]{(\ref{#1})}
\newcommand{\cH}{{\cal H}}
\newcommand{\cL}{{\cal L}}
\newcommand{\cP}{{\cal P}}
\newcommand{\Sloc}{{S_{\rm loc}}}
\newcommand{\cLloc}{{{\cal L}_{\rm loc}}}

\renewcommand{\bR}{\mbox{\boldmath $R$}}
\newcommand{\bS}{\mbox{\boldmath $S$}}
\newcommand{\bGam}{\mbox{\boldmath $\Gamma$}}
\newcommand{\bcH}{\mbox{\boldmath ${\cal H}$}}
\newcommand{\bcL}{\mbox{\boldmath ${\cal L}$}}

\renewcommand{\bZ}{\mbox{\boldmath $Z$}}

\newcommand{\hS}{\widehat{S}}
\newcommand{\hR}{\widehat{R}}
\newcommand{\hnab}{\widehat{\nabla}}
\newcommand{\cPt}{\widehat{\cP}}
\newcommand{\cHt}{\widehat{\cH}}
\newcommand{\tH}{\widetilde{H}}

\ifx\TwoupWrites\UnDeFiNeD\else\target{\magstepminus1}{11.3in}{8.27in}
\source{\magstep0}{7.5in}{11.69in}\fi
\newfont{\fourteencp}{cmcsc10 scaled\magstep2}
\newfont{\titlefont}{cmbx10 scaled\magstep3}
\newfont{\authorfont}{cmcsc10 scaled\magstep1}
\newfont{\fourteenmib}{cmmib10 scaled\magstep2}
\skewchar\fourteenmib='177
\newfont{\elevenmib}{cmmib10 scaled\magstephalf}
\skewchar\elevenmib='177
\makeatletter
\newcommand\nonsequentialeqnum{
\@addtoreset{equation}{section}
\def\theequation{\arabic{section}.\arabic{equation}}}
\newif\ifp@bblock \p@bblocktrue
\newcommand\nopubblock{\p@bblockfalse}
\newcommand\topspace{\hrule height 0pt depth 0pt \vskip}
\newcommand\p@bblock{\begingroup \tabskip=\hsize minus \hsize
\baselineskip=1.5\ht\strutbox \topspace-2\baselineskip
\halign to\hsize{\strut ##\hfil\tabskip=0pt\crcr
\the\Pubnum\crcr\the\date\crcr}\endgroup}

\renewcommand\titlepage{\ifx\TwoupWrites\UnDeFiNeD\null
\vspace{-1.7cm}\fi
\vskip0.6cm
\ifp@bblock\p@bblock \else\hrule height 0pt \relax \fi}
\makeatother
\newtoks\date
\newtoks\Pubnum
\newtoks\pubnum
\Pubnum={
~ \crcr
~ \crcr
YITP-01-18 \crcr
{\tt hep-th/0103187} \crcr
{}March 2001
}
\newcommand{\frontpageskip}{\vspace{12pt plus .5fil minus 2pt}}
\renewcommand{\title}[1]{\frontpageskip
\begin{center}{\titlefont #1}\end{center}\par}
\renewcommand{\author}[1]{\frontpageskip\par\begin{center}
{\authorfont #1}\end{center}
\nobreak
}

\renewcommand{\thanks}[1]{\footnote{#1}}
\renewcommand{\abstract}{\par\frontpageskip\centerline{
\fourteencp Abstract}
\vspace{8pt plus 3pt minus 3pt}}

\begin{document}

\begin{titlepage}

\vspace*{\fill}
\begin{center}
{\Large{\bf Higher-Derivative Gravity \\
and the AdS/CFT Correspondence 
}} \\
\vfill
{\sc Masafumi Fukuma
\footnote{E-mail: {fukuma@yukawa.kyoto-u.ac.jp}},
\sc So Matsuura
\footnote{E-mail: {matsu@yukawa.kyoto-u.ac.jp}}
and
{\sc Tadakatsu Sakai}
 \footnote{E-mail: {tsakai@yukawa.kyoto-u.ac.jp}}}\\[2em]

{\sl Yukawa Institute for Theoretical Physics, \\
      Kyoto University, Kyoto 606-8502, Japan } \\

\vfill
ABSTRACT
\end{center}
\begin{quote}

We investigate the AdS/CFT correspondence 
for higher-derivative gravity systems 
and develop a formalism in which 
the generating functional of the boundary field theory 
is given as a functional that depends only on 
the boundary values of bulk fields. 
We also derive a Hamilton-Jacobi-like equation 
that uniquely determines the generating functional, 
and give an algorithm calculating the Weyl anomaly. 
Using the expected duality between 
a higher-derivative gravity system 
and $N\!=\!2$ superconformal field theory in four dimensions, 
we demonstrate that the resulting Weyl anomaly 
is consistent with the field theoretic anomaly.

\end{quote}
\vfill
\end{titlepage}

%
\renewcommand{\thefootnote}{\arabic{footnote}}
\setcounter{footnote}{0}
\addtocounter{page}{1}%

\resection{Introduction}

Over the past few years, 
many attempts have been made to check the AdS/CFT correspondence 
\cite{M}\cite{GKP}\cite{W;holography}. 
(For a review, see Ref.\ \cite{review}). 
As an example, it is shown in Ref.\ \cite{W;holography} that 
the spectrum of chiral operators of ${\cal N}=4$ super Yang-Mills 
in four dimensions coincides with 
that of the Kaluza-Klein modes of type IIB supergravity on 
$AdS_5\times S^5$. 
Also, the computation of anomalies via bulk gravity has been shown 
to exactly reproduce the results of the super Yang-Mills theory 
\cite{W;holography}\cite{FMMR}\cite{HS;weyl}. 
However, this matching of the anomalies 
is valid only in the regime where 
$N\!\rightarrow\!\infty,~\lambda\!=\!g_{\rm YM}^2N\!\gg\! 1$, since 
the analysis is based on a classical supergravity 
computation. 
At present, it remains an important issue to test the duality 
beyond this regime.

There have been several attempts to confirm the validity 
of the duality beyond the classical gravity approximation 
\cite{AK}\cite{APTY}\cite{NO}\cite{BGN}\cite{BC}. 
Among these, Ref.\ \cite{APTY} treats 
${\cal N}\!=\!2~G\!=\!USp(N)$ superconformal field theory (SCFT) 
in four dimensions. 
This SCFT can be realized on the world volume of D$3$-branes situated
inside eight D$7$-branes coincident with an O$7^-$ brane, and 
is known \cite{N=2CFT} to be dual to type IIB string on 
$AdS_5\times S^5/\bZ_2$. 
The authors of Ref.\ \cite{APTY} showed that this duality reproduces 
the $1/N$ correction to the $U(1)_{\cal R}$ chiral anomaly correctly. 
In Refs.\ \cite{NO} and \cite{BGN}, 
the $1/N$ correction to the Weyl anomaly of 
the SCFT is computed using a higher-derivative gravity theory 
in which a curvature square term is added.

However, higher-derivative gravity theories\footnote{For a review of 
higher derivative gravity, see, e.g., Ref.\ \cite{HDG}. }
exhibit some features in the AdS/CFT correspondence 
that differ from those in Einstein gravity. 
To see this, we first recall that the equation of motion for Einstein 
gravity 
is a second-order differential equation in time $r$. 
Thus, a classical solution can be totally specified 
by prescribing the value at the boundary 
if we further impose the regular behavior of the solution 
inside the bulk \cite{W;holography}, 
and the boundary value can be identified with an external field 
coupled to an operator 
in the dual CFT \cite{GKP}\cite{W;holography}. 
The situation changes drastically 
if we consider higher-derivative theories. 
In fact, a higher-derivative system 
with Lagrangian density 
${\cal L}(g,\dot{g},\cdots,g^{(N\!+\!1)})$, 
where $g_{ij}$ is the metric and 
$\cdot=\partial/\partial r$, 
generically gives an equation of motion that is 
a differential equation of $2(N\!+\!1)$-order in $r$. 
We then would need $(N\!+\!1)$ boundary conditions 
for each field to specify a classical solution, 
even if we require its regular behavior inside the bulk.

The main aim of the present paper is to formulate 
higher-derivative gravity systems in accordance with the holographic 
principle. 
In this paper, we say that the holographic principle holds 
when the following two conditions are satisfied: 
(1)
~the classical solution of a higher-derivative system 
is specified uniquely by the boundary value of each bulk field, and  
(2)
~the bulk geometry becomes AdS-like near the boundary. 
In order to satisfy the first condition, 
we first note that the system 
${\cal L}(g,\dot{g},\cdots,g^{(N\!+\!1)})$ can be 
transformed into a Hamilton system 
with $(N\!+\!1)$ pairs of canonical variables 
$(g,Q^a),~(p,P_a)$ $(a=1,\cdots,N)$ by defining
$Q^a_{ij}=\partial^a g_{ij}/\partial r^a$. (See the next section for 
details.) 
Thus, by setting boundary conditions that are of the Dirichlet type 
for $g$ and the Neumann type for $Q^a$, 
the classical solution of this system can be specified 
only by the boundary value of $g$. 
Note also that the classical action of this system, 
which is obtained by plugging this solution into the action, 
becomes a functional of these boundary values of bulk fields. 
The second condition ensures the existence 
of a UV fixed point of the dual theory at the boundary, 
and such a fixed point enables us to take the continuum limit 
\cite{wilson-kogut}. 
We see below that appropriate boundary terms need to be added 
to the bulk action in order for the bulk metric 
to exhibit such asymptotic behavior when higher-derivative terms exist.

{}For a systematic treatment of these issues, 
we employ the Hamilton-Jacobi formulation, 
as introduced by de Boer, Verlinde and Verlinde \cite{dVV} 
to investigate the holographic RG structure of Einstein gravity. 
(See Refs.\ \cite{E.T.A}--\cite{DFGK} for more details of the holographic RG.) 
This formulation is further elaborated in Refs.\ \cite{vector}--\cite{KMM}. 
In particular, a systematic prescription for calculating the Weyl 
anomaly in arbitrary dimensions is developed in Ref.\ \cite{FMS}. 
In this paper, we show that the Hamilton-Jacobi equation is 
quite a useful tool also to study the holographic RG structure 
in higher-derivative systems. 
Actually, we can derive a Hamilton-Jacobi-like equation that 
determines the classical action in accordance with 
the holographic principle. 
That is, the classical action can be solved as a functional of a boundary 
value for each bulk field. 
As a check of our formulation, we compute $1/N$ corrections 
to the Weyl anomaly of the ${\cal N}\!=\!2$ SCFT 
by solving the Hamilton-Jacobi-like equation. 
In the course of this analysis, we find that the prescription 
developed in Ref.\ \cite{FMS} is again helpful. 
We show that our result can reproduce that of Refs.\ \cite{NO} and \cite{BGN}.

The organization of this paper is as follows. 
In \S 2, we formulate the Hamilton-Jacobi equation for 
a higher-derivative system with emphasis on applications to the AdS/CFT 
correspondence. 
In \S 3, we apply the formulation to higher-derivative gravity 
and derive an equation that determines the classical action. 
In \S 4, we solve the equation following the prescription given 
in Ref.\ \cite{FMS}, 
and demonstrate how to calculate the Weyl anomaly. 
We show that the resulting Weyl anomaly correctly reproduces 
that given in Refs.\ \cite{NO} and \cite{BGN}. 
Section 6 is devoted to a conclusion. 
There, a comment is given on the holographic RG structure in 
higher-derivative gravity systems. 
Some useful results are summarized in the appendices.

%
\resection{Hamilton-Jacobi equation for a  higher-derivative Lagrangian}

In this section, we give a prescription for determining the 
classical action when higher-derivative terms are added. 
We start our discussion for a system of point particles with the action 
\ba
 \bS[q(r)]=\int^t_{t^{\prime}} dr \,L\left(q,\dot{q},\cdots,q^{(N+1)}\right)
  \quad \left(q^{(n)}(r)\equiv d^nq(r)/dr^n\right)\,.
 \label{orig_action}
\ea
The extension of our argument to gravitational systems 
is straightforward and will be carried out in the next section.\footnote
{See also Ref.\ \cite{NK}, where higher-derivative systems are discussed 
from the viewpoint of string theories. 
}

The action \eq{orig_action} can be rewritten into the first-order form 
in the following way. 
We first introduce the Lagrange multipliers $p,P_1,\cdots,P_{N-1}$, so that 
$q,Q^1\!=\!\dot{q},\cdots,Q^N\!=\!q^{(N)}$ can be regarded as independent 
canonical variables: 
\ba
 &&\!\!\!\!\!\!\!\!\!\!\!\!\!\!\!
  L\left(q,Q^1,\cdots,Q^N,\dot{Q}^N;p,P_1,\cdots,P_{N-1}\right) \nn
 &&\!\!\!\!\!\!\!\!\!\!\!\!\!\!\!=
  p(\dot{q}-Q^1)+P_1(\dot{Q}^1-Q^2)+\cdots+P_{N-1}(\dot{Q}^{N-1}-Q^N) \nn
 &&\!\!\!\!\!\!\!\!\!
 +\,L(q,Q^1,\cdots,Q^N,\dot{Q}^N). 
\ea
We then carry out a Legendre transformation from $(Q^N,\dot{Q}^N)$ 
to $(Q^N,P_N)$ through 
\ba
 P_N={\partial L\over \partial\dot{Q}^N}
  \left(q,Q^1,\cdots,Q^N,\dot{Q}^N\right)\,. 
 \label{pn}
\ea
We here assume that this equation can be solved with respect to 
$\dot{Q}^N$ $
\left(\equiv f\!\left(q,Q^1,\cdots,Q^N;P_N\right)\right)$, 
and thus obtain the following action that is equivalent to \eq{orig_action} 
classically: 
\ba
 \bS[q,Q^1,\cdots,Q^N;\,p,P_1,\cdots,P_N]=\int^t_{t^\prime} dr \left[
  p\,\dot{q}+\sum_{a=1}^N P_a\dot{Q}^a-H(q,Q^a;\,p,P_a)\right], 
 \label{1st-action1}
\ea
where $\dot{Q}^N$ is now 
the time-derivative of the independent variable $Q^N$, 
and the Hamiltonian is given by
\ba
 H(q,Q^a;\,p,P_a)&=&p\,Q^1+P_1Q^2+\cdots + P_{N-1}Q^N
  +P_N \,f\!\left(q,Q^a;\,P_N\right) \nn
 &&-\,L\left(q,Q^1,\cdots,Q^N,f\!\left(q,Q^a;\,P_N\right)\right). 
 \label{1st-action2}
\ea
The variation of the action \eq{1st-action1} is given by 
\begin{align}
 \delta\bS=&\int^t_{t^\prime} dr \left[\,
  \delta p\left( \dot{q}-{\partial H\over\partial p}\right)
  +\sum_a\delta P_a\left(\dot{Q}^a-{\partial H\over\partial P_a}\right)
  \right. \nn 
  &\hspace{13mm}\left. 
  -\,\delta q
  \left( \dot{p}+{\partial H\over\partial q}\right)
  -\sum_a\delta Q^a\left(\dot{P}_a
  +{\partial H\over\partial Q^a}\right)\right] \nn
  &~~+\left( p\,\delta q+\sum P_a\,\delta Q^a\right)\Big|^t_{t^\prime}\, , 
\end{align}
and thus the equation of motion consists of the usual Hamilton equations, 
\ba
 \dot{q}={\partial H\over\partial p},\quad
  \dot{Q}^a={\partial H\over\partial P_a},\quad
  \dot{p}=-{\partial H\over\partial q},\quad 
  \dot{P_a}=-{\partial H\over\partial Q^a}\,,
 \label{hamiltoneq}
\ea
and the following constraint, which must hold at the boundary, 
$r=t$ and $r=t^\prime$: 
\ba 
 p\,\delta q+\sum_aP_a\,\delta Q^a =0   \quad (r=t, t')\,. 
 \label{bc}
\ea
The latter requirement, \eq{bc}, can be satisfied 
when we use either Dirichlet boundary conditions, 
\ba
 {\rm \underline{Dirichlet}:}\qquad \delta q=0\,,\quad \delta Q^a=0\quad 
(r=t,t')\,,
\ea
or Neumann boundary conditions, 
\ba
 {\rm \underline{Neumann}:}\qquad p=0\,,\quad P_a=0\quad (r=t,t')\,,
\ea
for each variable $q$ and $Q^a~(a\!=\!1,\cdots,N)$. 
If, for example, we take the classical solution $(\bar{q},\bar{Q}^a,
\bar{p},\bar{P}_a)$ that satisfies the Dirichlet boundary conditions 
for all $(q,Q^a)$ with the specified boundary values as 
\ba
 \bar{q}(r\!=\!t)=q,~\bar{Q}^a(r\!=\!t)=Q^a,\quad~{\rm and}\quad 
  \bar{q}(r\!=\!t^\prime)=q^\prime,~\bar{Q}^a(r\!=\!t^\prime)=Q^{\prime a}\,,
\ea
then after plugging the solution into the action, 
we obtain the classical action that is a function of these boundary values, 
\ba
 S(t,q,Q^a;\,t^\prime,q^\prime,Q^{\prime a})=
  \bS\left[\bar{q}(r),\bar{Q}^a(r);\,\bar{p}(r),\bar{P}_a(r)\right]. 
\ea
However, as we discussed in the Introduction, 
this classical action is not of great interest to us 
in the context of the AdS/CFT correspondence, 
since the holographic principle requires 
that the bulk be specified by only the values $q$ and $q'$ 
at the boundary. 
This leads us to use mixed boundary conditions: 
\ba
 \delta q=P_a=0\quad (r=t,\,t')\,.
\ea
That is, we impose Dirichlet boundary conditions for $q$ and 
Neumann boundary conditions for $Q^a$. 
In this case, the classical action 
(to be called the {\em reduced classical action}) 
becomes a function only of the boundary values $q$ and $q^\prime$: 
\ba
 S=S(t,q;\,t^\prime,q^\prime)\,. 
 \label{caction;toy}
\ea
A renormalization group interpretation of this condition 
is discussed briefly in the concluding section, and 
will be discussed in detail in a forthcoming paper \cite{FMS;fu}.

Now we derive a Hamilton-Jacobi-like equation that determines the 
reduced classical action (\ref{caction;toy}). 
This can be derived in two ways, 
and we start with the more complicated way, 
since this gives us a deeper understanding of the mathematical structure. 
To this end, we first change the
polarization of the system by performing the canonical 
transformation\footnote
{The following procedure corresponds to a change of representation 
from the $Q$-basis to the $P$-basis in the WKB approximation:
\ba
 \Psi(t,q,Q)=e^{i S(t,q,Q)/\hbar}\rightarrow
  \widehat{\Psi}(t,q,P)=e^{i \widehat{S}(t,q,P)/\hbar}\equiv
  \int dQ \,e^{-i P_a Q^a/\hbar}\,\Psi(t,q,Q)\,.\n
\ea
}
\ba
\widehat{\bS}\equiv\bS-\int^t_{t^\prime}dF\,, 
\ea
with the generating function
\ba
{}F=\sum_aP_aQ^a\,. 
\ea
Although the Hamilton equation does not change under this transformation, 
the boundary conditions at $r=t$ and $r=t^\prime$ become 
\ba
 p\,\delta q-\sum_a Q^a\delta P_a=0\quad(r=t,t')\,. 
\ea
These boundary conditions can be satisfied by imposing the Dirichlet 
boundary conditions for both $\bar{q}$ and $\bar{P}_a$: 
\ba
\bar{q}(r\!=\!t)=q,~\bar{P}_a(r\!=\!t)=P_a\,,\quad{\rm and}\quad 
\bar{q}(r\!=\!t^\prime)=q^\prime,~\bar{P}_a(r\!=\!t^\prime)=P^\prime_a\,. 
\ea
Substituting this solution into $\widehat{\bS}$, 
we obtain a new classical action that is a function 
of these boundary values, 
\ba
 \widehat{S}\left(t,q,P_a;\,t^\prime,q^\prime,P^\prime_a\right)
  =\widehat{\bS}\left[\bar{q}(r),\bar{Q}^a(r);\,
  \bar{p}(r),\bar{P}_a(r)\right]. 
\ea
By taking the variation of $\widehat{\bS}$ and using the equation of 
motion, we can easily show that the new classical action $\widehat{S}$ 
obeys the Hamilton-Jacobi equation: 
\ba
 {\partial\widehat{S}\over\partial t}&\!\!\!=\!\!\!&
  -H\left(q,-{\partial\widehat{S}\over\partial P_a};\,
 +{\partial\widehat{S}\over\partial q},P_a\right), \nn
  {\partial\widehat{S}\over\partial t^\prime}&\!\!\!=\!\!\!&
 +H\left(q^\prime,+{\partial\widehat{S}\over\partial P^\prime_a};\,
  -{\partial\widehat{S}\over\partial q^\prime},P^\prime_a\right). 
 \label{HJ;after}
\ea
The reduced classical action $S(t,q;t^\prime,q^\prime)$ is then obtained 
by setting $P_a\!=\!0$ in $\widehat{S}$: 
\ba
 S\left(t,q;t^\prime,q^\prime\right)=
  \widehat{S}\left(t,q,P_a\!=\!0;\,
  t^\prime,q^\prime,P^\prime_a\!=\!0\right). 
\ea
Note that the generating function $F$ vanishes at the boundary 
when we set $P_a\!=\!0$.

Here we briefly describe how the Hamilton-Jacobi equation (\ref{HJ;after}) 
is solved. 
{}For simplicity, we consider the case $N\!=\!1$ 
and focus only on the upper boundary at $r\!=\!t$. 
Motivated by the gravitational system 
considered in the next section, 
we assume that the Lagrangian takes the form 
\ba
L(q,\dot{q},\ddot{q})=L_0(q,\dot{q})+c L_1(q,\dot{q},\ddot{q}), 
\label{lag}
\ea
where 
\ba
 L_0(q,\dot{q})&\!\!\!=\!\!\!&{1\over 2}
  m_{ij}(q)\dot{q}^i\dot{q}^j-V(q), \nn
 L_1(q,\dot{q},\ddot{q})&\!\!\!=\!\!\!&
  {1\over 2}n_{ij}(q)\ddot{q}^i\ddot{q}^j
  -A_i(q,\dot{q})\ddot{q}^i-\phi(q,\dot{q}), 
 \label{lag1}
\ea
with
\ba
 A_i(q,\dot{q})&\!\!\!=\!\!\!&a_{ijk}^{(2)}(q)\dot{q}^j\dot{q}^k
  +a_{i}^{(0)}(q), \nn
 \phi(q,\dot{q})&\!\!\!=\!\!\!&
  \phi^{(4)}_{ijkl}(q)\dot{q}^i\dot{q}^j\dot{q}^k\dot{q}^l
  +\phi^{(2)}_{ij}(q)\dot{q}^i\dot{q}^j+\phi^{(0)}(q). 
 \label{lag2}
\ea
{}We further assume that the determinants of the 
matrices $m_{ij}(q)$ and $n_{ij}(q)$ have the same 
signature.\footnote
{In fact, it is easy to see that this is the 
case in the higher-derivative gravity system considered below.}
Following the procedure discussed above, 
this Lagrangian can be rewritten into the first-order form 
\ba
 L=p\,\dot{q}+P\dot{Q}-H(q,Q;\,p,P)\,,
\ea
with the Hamiltonian 
\begin{align}
 H(q,Q;p,P) =&\,\,p_iQ^i-{1\over 2}m_{ij}(q)Q^iQ^j+V(q) \nn
 &+{1\over 2c}
  n^{ij}(q)\Big(P_i+cA_i(q,Q)\Big)\Big(P_j+cA_j(q,Q)\Big)
  +c\,\phi(q,Q), 
 \label{trueH}
\end{align}
where $n^{ij}=(n_{ij})^{-1}$. 
The Hamilton-Jacobi equation \eq{HJ;after} is solved 
as a double expansion with respect to $c$ and $P$ 
by assuming that the classical action takes the form 
\ba
 \widehat{S}(t,q,P)&\!\!\!=\!\!\!&
  {1\over\sqrt{c}}\,\widehat{S}_{-1/2}(t,q,P)+\widehat{S}_0(t,q,P)
  +\sqrt{c}\,\widehat{S}_{1/2}(t,q,P)+c\,\widehat{S}_1(t,q,P) \nn
 &&+{\cal O}(c^{3/2}). 
 \label{step_first}
\ea
After some simple algebra, the coefficients are found to be 
\ba
 \widehat{S}_{-1/2}&\!\!\!=\!\!\!&
  {1\over 2}u^{ij}(q)P_iP_j+{\cal O}(P^3),\nn
  \widehat{S}_{0}&\!\!\!=\!\!\!&S_0(t,q)-P_i\,\partial^iS_0+{\cal O}(P^2),
\nn
 \widehat{S}_{1/2}&\!\!\!=\!\!\!&P_i\,u^{ij}(q)n_{jk}(q)
  \left[ \bGam^k_{lm}\,\partial^lS_0\,\partial^mS_0+\partial^kV(q)
  +n^{kl}(q)A_l\!\left( q,{\partial S_0\over\partial q}\right)\right] \nn
 &&+{\cal O}(P^2).
\ea
Here, 
\ba
\partial_i\equiv {\partial\over\partial q^i},\qquad 
\partial^i\equiv m^{ij}\partial_i, 
\ea
and $\bGam^i_{jk}$ is the affine connection defined by $m_{ij}$. 
Also $u^{ij}$ is defined by the relation 
\ba
u^{ik}(q)u^{jl}(q)m_{kl}(q)=n^{ij}(q). 
\ea
Furthermore, $S_{0}(t,q)=\widehat{S}_{0}(t,q,P\!=\!0)$ and 
$S_{1}(t,q)=\widehat{S}_{1}(t,q,P\!=\!0)$
satisfy the equations 
\begin{align} 
 -{\partial S_0\over\partial t} =&\, {1\over 2}m_{ij}(q)
  {\partial S_0\over\partial q^i}{\partial S_0\over\partial q^j}+V(q), \nn
 -{\partial S_1\over\partial t} =&\, m_{ij}(q)
  {\partial S_1\over\partial q^i}{\partial S_0\over\partial q^j} \nn
  &-{1\over 2}n_{ij}(q)
  \left(\bGam^i_{kl}\,\partial^kS_0\,\partial^lS_0+\partial^iV(q)\right)
  \left(\bGam^j_{mn}\,\partial^mS_0\,\partial^nS_0+\partial^jV(q)\right) \nn
  &-A_i\!\left(q,{\partial S_0\over\partial q}\right)
  \left(\bGam^i_{kl}\,\partial^kS_0\,\partial^lS_0+\partial^iV(q)\right)
  +\phi\!\left( q,{\partial S_0\over\partial q}\right)\,,
 \label{sol}
\end{align}
which can be expressed as a Hamilton-Jacobi-like equation 
for the reduced classical action 
$S(t,q)\!=\!S_0(t,q)+c\,S_1(t,q)+{\cal O}(c^2)$: 
\ba
 -{\partial S\over\partial t}=\tH(q,p),\qquad
  p_i={\partial S\over\partial q^i}\,,
 \label{HJ;c1}
\ea
where 
\ba
\tH(q,p)&\!\!\!=\!\!\!&{1\over 2}m^{ij}(q)p_ip_j+V(q) \nn
&&+\,c\left[ 
-{1\over 2}n_{ij}(q)
\left(\bGam^i_{kl}\,p^kp^l+\partial^iV(q)\right)
\left(\bGam^j_{mn}\,p^mp^n+\partial^jV(q)\right) \right.\nn
&&~~~~~~-
A_i(q,p)\left(\bGam^i_{kl}\,p^kp^l+\partial^iV(q)\right)
+\phi(q,p)\Bigl].
\label{HJ;c2}\label{step_last}
\ea
It is important to note that $\tH$ is not the Hamiltonian. 
In fact, the Hamilton equation for $\tH$ does not 
coincide with that obtained from (\ref{trueH}).

In solving the full Hamilton-Jacobi equation \eq{HJ;after} for 
$\widehat{S}(t,q,P)$, 
we imposed the condition that everything becomes regular 
around $c\!=\!0$ when we set $P\!=\!0$. 
This is because in most interesting cases 
(like those of the gravity systems we discuss in the following sections) 
the higher-derivative term is regarded as a perturbation, 
so that the reduced classical action must have a finite limit 
for $c\!\rightarrow\!0$. 
Once such a regularity condition is imposed, 
we have an alternative way to derive 
this {\em pseudo-Hamiltonian} $\tH$ with greater ease. 
In fact, for any Lagrangian of the form 
\ba
 L(q^i,\dot{q}^i,\ddot{q}^i)=
  L_0(q^i,\dot{q}^i)+c\,L_1(q^i,\dot{q}^i,\ddot{q}^i)\,,
\ea
one can prove the following theorem, 
assuming that the classical solution can be expanded 
around $c\!=\!0$:\footnote
{As long as we think of $L_1$ as a perturbation, 
any classical solution can be expanded as 
\ba
 \bar{q}(r)=\bar{q}_0(r)+c\,\bar{q}_1(r)+{\cal O}(c^2)\,.\n
\ea
Here $\bar{q}_0$ is the classical solution for $L_0$, and 
$\bar{q}_1$ is obtained by solving a second-order differential 
equation. 
Note that we can, in particular, enforce the boundary conditions 
\ba
 \bar{q}_0(r\!=\!t)=q,\quad \bar{q}_1(r\!=\!t)=0~~{\rm and}~~ 
  \bar{q}_0(r\!=\!t^{\prime})=q^{\prime},\quad \bar{q}_1(r\!=\!t^{\prime})=0\,.
 \label{bc;simple}\n
\ea
In this case, due to the equation of motion for $\bar{q}_0(r)$\,, 
the classical action is simply given by 
\ba
 S(q,t;q^{\prime},t^{\prime})
  =\int^t_{t^{\prime}} dr\big[ L_0(\bar{q}_0,\dot{\bar{q}}_0)
  +c\,L_1(\bar{q}_0,\dot{\bar{q}}_0,\ddot{\bar{q}}_0)\big]+{\cal O}(c^2)\,.
\n
\ea
This corresponds to the classical action considered in Ref.\ \cite{BGN}.
}

\noindent
\underline{\bf Theorem}\\
{\em Let $H_0(q,p)$ be the Hamiltonian corresponding to $L_0(q,\dot{q})$. 
Then the reduced classical action $S(t,q;\,t',q')
\!=\!S_0(t,q;\,t',q')+c\,S_1(t,q;\,t',q')+{\cal O}(c^2)$ 
satisfies the following equation up to ${\cal O}(c^2)$:}
\ba
 -{\partial S\over\partial t}=\tH(q,p), \quad 
  p_i={\partial S\over\partial q^i}, \quad{\rm and}\quad
  +{\partial S\over\partial t'}=\tH(q',p'), \quad 
  p'_i=-{\partial S\over\partial q^{\prime\,i}},
\label{theorem1}
\ea
{\em where} 
\ba
\tH(q,p)&\!\!\!\equiv\!\!\!&H_0(q,p)-c\,L_1(q,f_1(q,p),f_2(q,p)), \nn
f_1^i(q,p)&\!\!\!\equiv\!\!\!&\left\{H_0,q^i\right\}
  ={\partial H_0\over\partial p_i}, \nn
f_2^i(q,p)&\!\!\!\equiv\!\!\!&\left\{H_0,\left\{H_0,q^i\right\}\right\}
 ={\partial^2 H_0\over\partial p_i\partial q^j}
 {\partial H_0\over\partial p_j}
 -{\partial^2 H_0\over\partial p_i\partial p_j}
 {\partial H_0\over\partial q^j}.
\nn
 &&\left(\left\{F(q,p),G(q,p)\right\}\equiv
  \frac{\partial F}{\partial p_i}\frac{\partial G}{\partial q^i}-
  \frac{\partial G}{\partial p_i}\frac{\partial F}{\partial q^i}\right)
\label{theorem2}
\ea

\noindent
A proof of this theorem is given in Appendix A. 
It can easily be confirmed that this correctly reproduces 
(\ref{HJ;c1}) and (\ref{HJ;c2}) for the Lagrangian given in
\eq{lag}--\eq{lag2}.

%
\resection{Application to higher-derivative gravity}

In this section, following the prescription developed 
in the previous section, we derive an equation that determines 
the reduced classical action for a higher-derivative gravity system.

We first recall the holographic description of RG flows 
in the dual boundary field theory. 
We parametrize the bulk metric with the Euclidean ADM decomposition. 
(For more details of the ADM decomposition, see Appendix B.)
We then have
\ba
 ds^2&=&\widehat{g}_{\mu\nu}\,dX^\mu dX^\nu \nn
 &=&N(x,r)^2 dr^2+g_{ij}(x,r)\Bigl(dx^{i}+\lambda^i(x,r)dr\Bigr) 
  \Bigl(dx^{j}+\lambda^{j}(x,r)dr\Bigr).
\ea
Here $X^\mu=(x^i,r)$, with $i,j=1,2,\cdots,d$, 
and $N$ and $\lambda^i$ are the lapse and the shift function, 
respectively. 
The signature of the metric $g_{ij}$ is taken to be $(+\cdots +)$. 
By assuming that the geometry becomes AdS-like in the limit 
$r\!\rightarrow\!-\infty$, 
the Euclidean time $r$ is identified with 
the RG parameter of the $d$-dimensional boundary theory, 
and the time evolution of other bulk fields (such as scalars) 
is interpreted as an RG flow of the coupling constants 
with a UV fixed point at the boundary. 
To avoid a singularity of the metric $g_{ij}$ at $r\!=\!-\infty$, 
we restrict the region of $r$ such that $r_0\!\leq r\!<\!\infty$ 
\cite{GKP}\cite{W;holography}\cite{SW}. 
This corresponds to the introduction of a UV cut-off 
to the boundary field theory. 
In the following, we consider a $(d+1)$-dimensional manifold 
$M_{d+1}\!=\!\{(x^i,r)\}$ that has 
a topology given by $M_{d+1}\!\sim\! 
\left(\bR^d\cup\infty\right)\times\bR_+$, 
with $r_0\!\leq\!r\!<\! \infty$.

We consider classical gravity on $M_{d+1}$ with the action 
\ba
 \bS=\bS_B+\bS_b\,.
 \label{model}
\ea
Here $\bS_B$ is the bulk action given by 
\ba
 \bS_B&=&\int_{M_{d+1}} d^{d+1}X\sqrt{\widehat{g}}
  \,{\cal L}_B\,,\\
 {\cal L}_B&=&2\Lambda-\widehat{R}
  -a \widehat{R}^2-b\widehat{R}_{\mu\nu}^2
  -c\widehat{R}_{\mu\nu\rho\sigma}^2\,,
\ea
where $a,b$ and $c$ are some given constants. 
$\bS_b$ contains boundary terms defined on the boundary 
$\Sigma_d=\partial M_{d+1}$ at $r=r_0$. 
The form of $\bS_b$ can be determined by requiring that 
it is invariant under the diffeomorphism 
\ba
 X^{\mu}\rightarrow X^{\prime\mu}=f^{\mu}(X), 
\ea
with the condition 
\ba
 f^r (r=r_0,x)=r_0. 
 \label{diffeo;boundary}
\ea
Equation \eq{diffeo;boundary} implies that the diffeomorphism does not change 
the location of the boundary. 
It is then easy to verify that $\bS_b$ takes the form (for details see 
Appendix C) 
\ba
\bS_b&\!\!\!=\!\!\!&\int_{\Sigma_d}d^dx\sqrt{g}\,{\cal B},
\label{Sb}
\ea
with
\ba
 {\cal B}=2K+x_1\,RK+x_2\,R_{ij}K^{ij}+x_3\,K^3
  +x_4\,KK_{ij}^2+x_5\,K_{ij}^3,
\ea
where $K_{ij}$ is the extrinsic curvature of $\Sigma_d$ given by 
\ba
 K_{ij}={1\over 2N}\left(\dot{g}_{ij}
  -\nabla_i\lambda_j-\nabla_j\lambda_i\right),
\ea
and $K=g^{ij}K_{ij}$. 
$\nabla_i$ and $R_{ijkl}$ are, respectively, the covariant derivative 
and the Riemann tensor defined by $g_{ij}$. 
The first term in ${\cal B}$ ensures that 
the Dirichlet boundary conditions can be imposed 
in the Einstein theory \cite{GH} 
and also plays an important role in the context of the 
AdS/CFT correspondence \cite{BK2}. 
We argue below that the coefficients $x_1,\cdots,x_5$ 
must obey some relations so that the holography holds 
even for higher-derivative gravity.\footnote
{See, e.g., Refs.\ \cite{Mye} and \cite{NO;boundary} for another discussion 
of boundary terms in higher-derivative gravity. 
}

The action (\ref{model}) is expressed in terms of the ADM parametrization as 
\ba
 \bS&=&\int_{M_{d+1}}d^{d+1}X \left[\sqrt{\hg}\,
  \cL_B-\frac{\partial}{\partial r}
  \left(\sqrt{g}\,{\cal B}\right)\right]\nn
 &=&\int_{r_0}^\infty dr \int d^d x \sqrt{g}\,
  \left[\bcL_0(g,K;\,N,\lambda)
  +\bcL_1(g,K,\dot{K};\,N,\lambda)
  \right],
 \label{action}
\ea
where\footnote
{We here use the following abbreviated notation: 
 $K^n_{ij}\equiv K_{i_1}^{i_2}K_{i_2}^{i_3}\cdots K_{i_n}^{i_1},\,
 (K^2)_{ij}\equiv K_{ik}K^k_j$.
}
\ba
 \frac{1}{N}\,\bcL_0&\!\!\!=\!\!\!&2\Lambda-R
+K_{ij}^2-K^2, 
\ea
\ba
 \frac{1}{N}\,\bcL_1&\!\!\!=\!\!\!&-aR^2-bR_{ij}^2-cR_{ijkl}^2 
+\Big[(-6a+2x_1)K_{ij}^2+(2a-x_1)K^2\Big]R \nn
&&+\Big[-2(2b+4c-x_2)(K^2)_{ij}+(2b+2x_1-x_2)KK_{ij}\Big]R^{ij} \nn
&&+\,2(6c+x_2)K_{ik}K_{jl}R^{ijkl} \nn
&&-\,2(2b+c-3x_5)K_{ij}^4+(4b+4x_4-x_5)KK_{ij}^3 \nn
&&-\,(9a+b+2c-2x_4)\left(K_{ij}^2\right)^2
+(6a-b+6x_3-x_4)K^2K_{ij}^2 \nn
&&-\,(a+x_3)K^4 \nn 
&&-\,(4b+2x_1-x_2)K_{ij}\nabla^i\nabla^jK
+2(b-4c+x_2)K_{ij}\nabla^j\nabla_kK^{ki} \nonumber
\ea
\ba
&&+\,(8c+x_2)K_{ij}\nabla^2K^{ij}+2(b+x_1)K\nabla^2K \nn
&&-\Big[(4a+b)g^{ij}g^{kl}+(b+4c)g^{ik}g^{jl}\Big]L_{ij}L_{kl} \nn
&&+\bigg[\Big\{(4a-x_1)R+(12a+2b-x_4)K_{kl}^2-(4a+3x_3)K^2\Big\}g^{ij} 
\nn
&&~~~+(2b-x_2)R^{ij}+(4b+8c-3x_5)(K^2)^{ij}
-2(b+x_4)KK^{ij}\bigg]L_{ij}, 
\ea
with
\ba
 K_{ij}&=&\frac{1}{2N}
  \left(\dot{g}_{ij}-\nabla_i\lambda_j-\nabla_j\lambda_i\right)\,,
 \label{kij}\\
 L_{ij}&=&\frac{1}{N}\left( \dot{K}_{ij}
  -\lambda^k\,\nabla_kK_{ij}
  -\nabla_i\lambda^k\,K_{kj}
  -\nabla_{j}\lambda^k\,K_{ik}
  +\nabla_i\nabla_{j}N\right)\,. 
 \label{lij}
\ea
By regarding $g_{ij}$ and $K_{ij}$ as independent canonical 
variables,\footnote
{The correspondences between the variables in \S2 are
as follows: 
$q\leftrightarrow g_{ij},\,p\leftrightarrow \sqrt{g}\,\pi^{ij},\,
Q\leftrightarrow K_{ij},\,P\leftrightarrow \sqrt{g}\,P^{ij}$.
}
the action \eq{action} can be further rewritten into the first-order form 
\ba
 \bS&\!\!\!=\!\!\!&\int_{r_0}^\infty dr \int d^d x \sqrt{g}\left[ 
  \pi^{ij}\left( 
  \dot{g}_{ij}-2NK_{ij}-\nabla_i\lambda_j-\nabla_j\lambda_i\right)
  +\bcL_0+\bcL_1\right] \nn
 &\!\!\!=\!\!\!&\int_{r_0}^\infty dr \int d^d x \sqrt{g}\left[ 
  \pi^{ij}\dot{g}_{ij}+P^{ij}\dot{K}_{ij}-\bcH(g,K;\pi,P;N,\lambda) \right]
\,.
 \label{action2}
\ea
Here the Hamiltonian density $\bcH$ can be evaluated as
\ba
 \bcH&\!\!\!=\!\!\!&\pi^{ij}
  \left(2NK_{ij}+\nabla_i\lambda_j+\nabla_j\lambda_i\right)
  +P^{ij}\dot{K}_{ij}-\bcL_0 -\bcL_1\nn
 &\!\!\!=\!\!\!&N\cH(g,K;\pi,P)+\lambda_i\cP^i(g,K;\pi,P), 
\ea
with
\ba
\cH(g,K;\pi,P)&\!\!\!=\!\!\!&2\pi^{ij}K_{ij}-{1\over 4(b+4c)}P_{ij}^2
+{4a+b\over 4(b+4c)\left(4da+(d+1)b+4c\right)}P^2 \nn
&&-\,\nabla_i\nabla_jP^{ij}
+\left[ A_1\,R^{ij}+A_2\,(K^2)^{ij}+A_3\,KK^{ij}\right]P_{ij} \nn
&&+\left[A_4\,R+A_5\,K_{ij}^2+A_6\,K^2\right]P \nn
&&-\,2\Lambda+R-K_{ij}^2+K^2 \nn
&&+\,B_1\,R^2+B_2\,R_{ij}^2+B_3\,R_{ijkl}^2 \nonumber
\ea
\ba
&&+\,\left(C_1\,K_{ij}^2+C_2\,K^2\right)R
+\left[C_3\,(K^2)_{ij}+C_4KK_{ij}\right]R^{ij} \nn
&&+\,C_5\,K_{ik}K_{jl}R^{ijkl} \nn
&&+\,D_1\,K_{ij}^4+D_2\,KK_{ij}^3+D_3\, (K_{ij}^2)^2 
+D_4\,K^2K_{ij}^2 
+D_5\,K^4 \nn
&&+\,E_1\,K_{ij}\nabla^i\nabla^jK+E_2\,K_{ij}\nabla^j\nabla_kK^{kj} \nn
&&+\,E_3\,K_{ij}\nabla^2K^{ij}+E_4\,K\nabla^2K, \\
\cP^i(g,K;\pi,P)&\!\!\!=\!\!\!&
-\,2\nabla_j\pi^{ij}+P_{kl}\nabla^iK^{kl}-2\nabla^k(K^{ij}P_{jk}). 
 \label{ham}
\ea
The coefficients $A_1,\cdots,E_4$ are not important 
in the following discussion, and are listed in Appendix D. 
The classical equivalence between the two actions \eq{action} 
and \eq{action2} can be easily established by noting that 
the latter gives the following equation of motion for $\pi^{ij}$: 
\ba
 P^{ij}
  &\!\!\!=\!\!\!&-\,2\left((4a+b)\, g^{ij}g^{kl}+(b+4c)\, 
  g^{ik}g^{jl}\right)L_{kl}\nn
 &&+\left[ (4a-x_1)R+(12a+2b-x_4)K_{kl}^2-(4a+3x_3)K^2\right]g^{ij} \nn 
 &&+\,(2b-x_2)R^{ij}+(4b+8c-3x_5)(K^2)^{ij}-2(b+x_4)\,KK^{ij}\,.
 \label{PKdot}
\ea 
This correctly reproduces the original action \eq{action} 
when substituted into \eq{action2}.

{}Following the prescription given in \S $2$, 
we now make a canonical transformation that 
changes  the polarization of $\bS$ from $(g_{ij},K_{ij})$ 
to $(g_{ij},P^{ij})$: 
\ba
\widehat{\bS}&\!\!\!\equiv\!\!\!&\bS-\int_{M_{d+1}}d^{d+1}X\,
{\partial\over\partial r}
\left( \sqrt{g}\,K_{ij}P^{ij}\right) \nn
&\!\!\!=\!\!\!&\int_{r_0}^\infty \int d^d x\,\sqrt{g}
\left( 
\pi^{ij}\dot{g}_{ij}-K_{ij}\dot{P}^{ij}-N\widehat{\cH}
-\lambda_i\widehat{\cP}^i\right), 
\ea
with 
\ba
 \cHt(g,K;\pi,P)&\!\!\!\equiv\!\!\!&\cH(g,K;\pi,P)+KK_{ij}P^{ij}, \nn
 \cPt^i(g,K;\pi,P)&\!\!\!\equiv\!\!\!&
  \cP^i(g,K;\pi,P)-\nabla^i(K_{jk}P^{jk}) \nn 
 &\!\!\!=\!\!\!&-2\nabla_j\pi^{ij}-\nabla^iP^{jk}\,K_{jk}
  -2\nabla^k(K^{ij}P_{jk}), 
\ea
where we have used the relation 
\ba
 \partial_r\sqrt{g}=\sqrt{g}\left( NK+\nabla^i\lambda_i\right). 
\ea
Since $N$ and $\lambda_i$ are the Lagrange multipliers, 
we obtain the Hamiltonian and momentum constraints 
\begin{eqnarray}
 {1 \over \sqrt{g}}\,{\delta \widehat{\bS} \over \delta N}
  &=&\cHt(g,K;\,\pi,P)~=~0,
 \label{h;constraint}\\
 {1 \over \sqrt{g}}\,{\delta \widehat{\bS} \over \delta \lambda_{i}}
  &=&\cPt^{i}(g,K;\,\pi,P)~=~0.
 \label{m;constraint}
\end{eqnarray}

We now let $\bar{g}_{ij}$ and $\bar{P}^{ij}$ represent the solution 
to the equation of motion for $\widehat{\bS}$ 
that obeys the boundary conditions 
\ba
 \bar{g}_{ij}(x,r\!=\!r_0)=g_{ij}(x),\quad 
  \bar{P}^{ij}(x,r\!=\!r_0)=P^{ij}(x). 
\ea
We also require that the solution be regular or be set to 
some specific value inside the bulk ($r\!\rightarrow\!\infty$), 
and assume that the above boundary condition is sufficient 
to specify the classical solution completely 
\cite{W;holography}. 
Plugging the solution into $\widehat{\bS}$, we obtain the classical 
action $\hS[g(x),P(x)]$, 
which satisfies the following Hamilton-Jacobi equation:\footnote
{The last equation demonstrates the invariance of $\hS$ 
under a $d$-dimensional diffeomorphism, 
\ba
 0&\!\!\!=\!\!\!&-\int_{\Sigma_d} d^dx\sqrt{g}\,\epsilon_i\cPt^i \nn
 &\!\!\!=\!\!\!&\int_{\Sigma_d} d^dx\left[ 
  (\nabla_i\epsilon_j+\nabla_j\epsilon_i){\delta \hS\over\delta g_{ij}}
  +(-\partial_k\epsilon^i\,P^{kj}-\partial_k\epsilon^j\,P^{ik}
  +\epsilon^k\,\partial_kP^{ij})
  {\delta \hS\over \delta P^{ij}}\right]\,, \n
\ea
with $\epsilon^i(x)$ an arbitrary function. 
This also demonstrates the invariance of the reduced classical action, 
\ba
 0=\int_{\Sigma_d} d^dx  
  \left(\nabla_i\epsilon_j+\nabla_j\epsilon_i\right)
  {\delta S\over\delta g_{ij}}\,, \n
\ea
for arbitrary $\epsilon^i(x)$. 
}
\ba
 {1\over\sqrt{g}}{\delta \hS\over\delta g_{ij}}=-\pi^{ij}\,,\quad
  {1\over\sqrt{g}}{\delta \hS\over\delta P^{ij}}=+K_{ij}\,, 
\ea
\ba
 \widehat{\cH}(g,K;\pi,P)&=&0 \label{h-const}\,,\\
 \widehat{\cP}^i(g,K;\pi,P)&=&0 \label{m-const}
 \label{HJ;gravity}\,.
\ea
Since the Hamiltonian density is a linear combination of the constraints, 
the classical action $\hS$ does not depend on the coordinate 
of the lower boundary: 
\ba
 {\partial\over\partial r_0}\hS 
  =\int d^d x \sqrt{g}\left(N\cHt+\lambda_i\cPt^i\right)=0\,. \label{eqa}
\ea
This implies that the reduced classical action 
\ba
 S[g(x)]\equiv \widehat{S}[g(x),P(x)\!=\!0]
\ea
is also independent of $r_0$: 
\ba
  {\partial\over\partial r_0}S=0. \label{eqA}
\ea

The Hamiltonian and the momentum constraints \eq{h-const} and \eq{m-const} 
can be translated into equations for the reduced classical action, 
as we sketched for point-particle systems 
in Eqs.\ \eq{step_first}--\eq{step_last}.
However, the resulting equation can be derived most easily by using 
the Theorem, (\ref{theorem1}) and (\ref{theorem2}), as follows: 
We first rewrite the Lagrangian density of zero-th order, $\bcL_0$, 
into the first-order form 
\ba
 \bcL_0 \rightarrow 
  \pi^{ij}\dot{g}_{ij}-\bcH_0\,, 
\ea
where the zero-th order Hamiltonian density $\bcH_0$ is given by 
\ba
 \bcH_0(g,\pi;\,N,\lambda)
  =N\,\left(\pi_{ij}^2-{1\over d-1}\pi^2-2\Lambda+R\right) 
  -2\lambda_i\,\nabla_j\pi^{ij}\,. 
\ea
Then by using the Theorem, the pseudo-Hamiltonian density 
is given by 
\ba
 \widetilde{\bcH}(g,\pi;\,N,\lambda)
  =\bcH_0(g,\pi;\,N,\lambda)-\bcL_1(g,K^0(g,\pi),K^1(g,\pi);\,N,\lambda)\,.
 \label{hj-like}
\ea
Here $K^0_{ij}(g,\pi)$ is obtained by replacing 
$\dot{g}_{ij}(x)$ in \eq{kij} with 
$\left\{\int d^d y \sqrt{g}\,\bcH_0(y),\,g_{ij}(x)\right\}$, 
and it is calculated to be 
\ba
  K^0_{ij}=\pi_{ij}-{1\over d-1}\pi\,g_{ij}\,. 
\ea
On the other hand, 
$K^1_{ij}\equiv\left\{\int d^d y \sqrt{g}\,\bcH_0(y),\,K^0_{ij}\right\}$ 
is found to be equivalent to replacing $L_{ij}$ in $\bcL_1$ 
by 
\begin{align}
 L^0_{ij} =&\, -{1\over 2(d-1)^2}\Bigl[ 
  2(d-1)\Lambda+(d-1)R+(d-1)\pi_{kl}^2-3\pi^2\Bigr] g_{ij} \nn 
  &~+R_{ij}+2(\pi^2)_{ij}-{3\over d-1}\pi\pi_{ij}\,.
 \label{eqlast}
\end{align}
Using Eqs.\ \eq{eqA}--\eq{eqlast}, 
we obtain the following Hamilton-Jacobi-like equation 
for the reduced classical action: 
\ba
 0&\!\!=\!\!&\int d^d x \sqrt{g}\,
  \widetilde{\bcH}\left(g(x),\pi(x);N,\lambda^i\right) \nn
 &\!\!=\!\!&\int d^d x \sqrt{g}\,\left[N\,\widetilde{\cH}(g(x),\pi(x))
  +\lambda^i\,\widetilde{\cP}_i(g(x),\pi(x))\right]\,,
 \label{hj-like2a}
\ea
\ba
 \pi^{ij}(x) =\frac{-1}{\sqrt{g}}\frac{\delta S}{\delta g_{ij}(x)}\,, 
 \label{hj-like2b}
\ea
where\footnote
{We have ignored those terms in $\widetilde{\cH}$ 
that contain the covariant derivative $\nabla$. 
This is justified when we consider the holographic Weyl anomaly 
in four dimensions. 
Actually, it turns out that they give only total derivative terms 
in the Weyl anomaly. 
}
\ba
 \widetilde{\cH}(g,\pi)
  &\equiv&\pi_{ij}^2-{1\over d-1}\pi^2-2\Lambda+R\nn
 &&+\,\alpha_1\,\pi_{ij}^4+\alpha_2\,\pi\pi_{ij}^3
  +\alpha_3\left(\pi_{ij}^2\right)^2+\alpha_4\,\pi^2\pi_{ij}^2
  +\alpha_5\,\pi^4 \nn
 &&+\,\beta_1\,\Lambda\pi_{ij}^2+\beta_2\,\Lambda\pi^2+\beta_3\,R\,\pi_{ij}^2
  +\beta_4\,R\,\pi^2\nn
 &&+\,\beta_5\,R_{ij}(\pi^2)^{ij}
  +\beta_6\,R_{ij}\,\pi\pi^{ij}
  +\beta_7\,R_{ijkl}\,\pi^{ik}\pi^{jl} \nn
 &&+\,\gamma_1\,\Lambda^2+\gamma_2\,\Lambda R+\gamma_3\,R^2+\gamma_4\,R_{ij}^2
  +\gamma_5\,R_{ijkl}^2\,,
 \label{hj-like2c}\\
 \widetilde{\cP}_i(g,\pi)&\equiv&-2\nabla^j\pi_{ij}\,,
 \label{hj-like2d}
\ea
with 
\begin{align}
\alpha_1 = &\, 2c, \quad \alpha_2={2x_5 \over (d-1)}, \nn
\alpha_3 =&\, {1\over 4(d-1)^2}
\Big[ 4a+(d^2-3d+4)b+4(d-2)(2d-3)c \nn
&\qquad\qquad~~ -2(d-1)(dx_4+3x_5) \Big], \nn
\alpha_4 =&\, {1\over 2(d-1)^3}\left[ 
-4a-(d^2-3d+4)b-4(2d^2-5d+4)c \right. \nn
&\left.\qquad\qquad~~ -3dx_3+(2d^2-7d+2)x_4-3(2d-1)x_5\right],\nn
\alpha_5 =&\, {1\over 4(d-1)^4}\left[ 
4a+(d^2-3d+4)b+4(2d^2-5d+4)c \right. \nn
&\left.\qquad\qquad~~ +2(3d-4)x_3-2(d^2-6d+6)x_4+2(5d-6)x_5\right], 
\end{align}
\begin{align}
\beta_1= &\, {1\over (d-1)^2}\Big[ 4da-d(d-3)b-4(d-2)c 
-(d-1)(dx_4+3x_5)\Big],\nn
\beta_2 =&\, {1\over (d-1)^3}\Big[-4da+d(d-3)b+4(d-2)c \nn
&\qquad\qquad~~-3dx_3+(d^2-2d-2)x_4+3(d-2)x_5 \Big], \nn
\beta_3 =&\, {1\over 2(d-1)^2}\Big[4a+(d^2-3d+4)b-4(3d-4)c \nn
&\qquad\qquad~~-(d-1)(dx_1+x_2-(d-2)x_4+3x_5)\Big], \nn
\beta_4 =&\, {1\over 2(d-1)^3}\Big[-4a-(d^2-3d+4)b+4(d-2)c \nn 
&\qquad\qquad~~-(d-1)(d-4)x_1-3(d-1)x_2+3(d-2)x_3\nn
&\qquad\qquad~~-(d^2-8d+10)x_4 +3(3d-4)x_5\Big], \nn
\beta_5 =&\, 16c+3x_5,\qquad 
\beta_6={2(x_1+2x_2-x_4-3x_5)\over d-1},\qquad
\beta_7=-12c-2x_2,
\end{align}
\ba
\gamma_1&\!\!\!=\!\!\!&{d\over (d-1)^2}\Big[ 4da+(d+1)b+4c \Big],\nn
\gamma_2&\!\!\!=\!\!\!&{1\over (d-1)^2}\Big[ 4da-d(d-3)b-4(d-2)c 
-(d-1)(dx_1+x_2)\Big],\nn
\gamma_3&\!\!\!=\!\!\!&{1\over 4(d-1)^2}\Big[ 4a+(d^2-3d+4)b-4(3d-4)c 
+2(d-1)((d-2)x_1-x_2)\Big],\nn
\gamma_4&\!\!\!=\!\!\!&4c+x_2,\qquad \gamma_5=c. 
\ea
Since the classical action $\hS[g(x),P(x)]$ 
is independent of the choice of $N$ and $\lambda^i$ 
(and, thus, so is $S[g(x)]$), 
from Eqs.\ \eq{hj-like2a}--\eq{hj-like2d} 
we finally obtain the following equation that determines the reduced 
classical action: 
\ba
 \widetilde{\cH}(g_{ij}(x),\pi^{ij}(x))=0\,,\quad 
  \widetilde{\cP}_i(g_{ij}(x),\pi^{ij}(x))=0\,,\quad 
  \pi^{ij}(x)=\frac{-1}{\sqrt{g}}\frac{\delta S}{\delta g_{ij}(x)}\,.
 \label{basic}
\ea

We conclude this section by making a few comments 
on the possible form of the boundary action $\bS_b$ 
and the cosmological constant $\Lambda$. 
As discussed above, in order that the boundary field theory 
has a continuum limit, 
the geometry must be asymptotically AdS: 
\ba
 ds^2\rightarrow dr^2+e^{-2r/l}\gamma_{ij}(x)dx^idx^j
\quad {\rm for}~~r\rightarrow -\infty. 
\label{asymp;ads}
\ea
This should be consistent with our boundary condition 
$P^{ij}\!=\!0$. 
By investigating the equation of motion derived from the action 
\eq{action2} explicitly, 
it can easily be shown that this compatibility gives rise to the relation 
\ba
 x_1&\!\!\!=\!\!\!&4a, \nn
 x_2&\!\!\!=\!\!\!&2b, \nn
 d^2\,x_3+d\,x_4+x_5&\!\!\!=\!\!\!&-{4\over 3}\Big( d(d+1)a+d\,b+2c\Big). 
 \label{coef}
\ea
It can also be shown that the asymptotic behavior (\ref{asymp;ads}) determines 
the cosmological constant $\Lambda$ as
\ba
 \Lambda=-{d(d-1)\over 2l^2}+{d(d-3)\over2l^4}
 \Big[ d(d+1)a+db+2c \Big]. 
 \label{cc}
\ea

%
\resection{Solution to the flow equation and the Weyl anomaly}

In this section, we solve the equation \eq{basic}, 
using the derivative expansion that was developed in Ref.\ \cite{FMS}. 
We then apply the result to computing the holographic Weyl 
anomaly of ${\cal N}\!=\!2$ superconformal field theory in four dimensions, 
which is dual to IIB supergravity on $AdS_5\times S^5/\bZ_2$.

We first note that the basic equation, \eq{basic}, 
can be rewritten as a flow equation of the form 
\ba
 \{ S, S \}+\{ S,S,S,S \} ={\cal L}_d, 
 \label{hj}
\ea
with 
\begin{align}
 \left( \sqrt{g}\right)^2\,\{ S, S \} \equiv &\, \left[  
  \left({\delta S\over\delta g_{ij}}\right)^2
  -{1\over d-1}\left( g_{ij}{\delta S\over \delta g_{ij}} \right)^2 
  \right. \nn
 &\left.
  ~+ \beta_1\,\Lambda \left( {\delta S \over \delta g_{ij}} \right)^2
  +\beta_2\,\Lambda \left( g_{ij}{ \delta S \over \delta g_{ij}} \right)^2
  +\beta_3\, R \left( { \delta S \over \delta g_{ij}} \right)^2
\right.\nn
&\left.
  ~+ \beta_4\, R \left( g_{ij}{ \delta S \over \delta g_{ij}} \right)^2
  +\beta_5\, R_{ij}g_{kl} 
   { \delta S \over \delta g_{ik}}{ \delta S \over \delta g_{jl}} 
\right.\nn
&\left.
  ~+ \beta_6\, R_{ij} { \delta S \over \delta g_{ij}}
  \,g_{kl}{ \delta S \over \delta g_{kl}} 
  +\beta_7\, R_{ijkl} { \delta S \over \delta g_{ik}}
  \,{ \delta S \over \delta g_{jl}} \right], 
\end{align}
\ba
 \left( \sqrt{g}\right)^4 \{ S,S,S,S \}&\!\!\!\equiv\!\!\!&\left[ 
  \alpha_1 \left({\delta S\over\delta g_{ij}}\right)^4
  +\alpha_2 \left(g_{kl}{\delta S\over\delta g_{kl}}\right)
  \left({\delta S\over\delta g_{ij}}\right)^3
  +\alpha_3 \left(\left({\delta S\over\delta g_{ij}}\right)^2\right)^2 
  \right.\nn
 &&\left. ~+ \alpha_4 
   \left(g_{kl}{\delta S\over\delta g_{kl}}\right)^2
   \left({\delta S\over\delta g_{ij}}\right)^2
  +\alpha_5 \left(g_{ij}{\delta S\over\delta g_{ij}}\right)^4 \right], 
\ea
\ba
 {\cal L}_d&\!\!\!\equiv\!\!\!&2\Lambda-R-\gamma_1\Lambda^2-\gamma_2\Lambda 
R
  -\gamma_3 R^2-\gamma_4R_{ij}^2-\gamma_5R_{ijkl}^2. 
\ea
{}Following Refs.\ \cite{dVV} and \cite{FMS}, 
we then assume that the reduced classical action $S[g(x)]$ takes the form
\begin{equation}
 \frac{1}{2\kappa_{d+1}^2}S[g(x)]
  =\frac{1}{2\kappa_{d+1}^2}S_{\rm loc}[g(x)]+\Gamma[g(x)]\,,
 \label{eh;on-shell}
\end{equation}
where $2\kappa_{d+1}^2$ is the $(d+1)$-dimensional Newton constant. 
The functional $\Gamma[g]$ is identified with the generating functional 
of the boundary field theory in the background metric $g_{ij}(x)$, 
with any local sources set to zero, 
and $\Sloc[g]$ is the local counterterm in $S[g]$: 
\begin{eqnarray}
  S_{\rm loc}[g(x)]
  &=&\int d^d x \,\sqrt{g(x)}\,\cLloc(x) \nn
  &=&\int d^d x\,
  \sqrt{g(x)}\sum_{w=0,2,4,\cdots}\bigl[\cLloc(x)\bigr]_w.
\end{eqnarray}
Here we have arranged the sum over local terms 
according to the weight $w$ \cite{FMS}, which is defined additively 
from the following rule:
\begin{center}
\begin{tabular}{c|c}
       & weight \\ \hline
$g_{ij}(x), \,\Gamma[g]$ & $0$ \\ \hline
$\partial_{i}$ & $1$ \\ \hline
$R, \,R_{ij},
\cdots$ & $2$ \\ \hline
$\delta \Gamma / \delta g_{ij}(x)$
& $d$ 
\end{tabular}
\end{center}
We then substitute \eq{eh;on-shell} into the flow equation (\ref{hj}) 
and rearrange the resulting equation with respect to the weight. 
The parts of weight $0$ and $2$ give  
\begin{eqnarray}
 2\Lambda-\gamma_1\Lambda^2&\!\!\!=\!\!\!&
  \Bigl[\left\{\Sloc,\,\Sloc\right\}\Bigr]_0
  +\Bigl[\{\Sloc,\,\Sloc,\Sloc,\Sloc\}\Bigr]_0
 \label{eqA0}\,,\\
 -R-\gamma_2\Lambda R&\!\!\!=\!\!\!&
  \Bigl[\left\{\Sloc,\,\Sloc\right\}\Bigr]_2
  +\Bigl[\{\Sloc,\,\Sloc,\Sloc,\Sloc\}\Bigr]_2
 \label{eqA2}\, .
\end{eqnarray}
These two equations determine $[\cLloc]_0$ and $[\cLloc]_2$ as 
\ba
  \left[\cLloc\right]_0= W\,, \qquad
  \left[\cLloc\right]_2= -\Phi\,R,
\ea
\begin{align}
 W =&\, -{2(d-1)\over l}+{1\over l^3}\Big[
  -4d(d+1)a-4db-8c+d(d^2x_3+dx_4+x_5)\Big], \nn
 \Phi =&\, {l\over d-2}-
  {2\over (d-1)(d-2)\,l}\Big[ d(d+1)a+d\,b+2c\Big]\nn
 &+{1\over l}\left[d\,x_1+x_2
  +{3(d^2x_3+d\,x_4+x_5)\over 2(d-1)}\right], 
 \label{wphi}
\end{align}
where (\ref{cc}) has been used. 
It is worthwhile to note that $W$ and $\Phi$ 
can be written in terms of only $a,b$ and $c$ upon substituting into 
(\ref{coef}): 
\ba
 W&\!\!\!=\!\!\!&-{2(d-1)\over l}-{4(d+3)\over 3l^3}\Big[
  d(d+1)a+db+2c\Big], \nn
 \Phi&\!\!\!=\!\!\!&{l\over d-2}
  +{2\over (d-2)\,l}
  \Big[ d(d-5)a-2b-2c\Big]. 
\ea
{}For $d\!>\!4$, the flow equation of weight $4$ simply determines 
$\bigl[\cLloc\bigr]_4$ in the local counterterm, 
as in the case of Einstein gravity (cf.\ Ref.\ \cite{FMS}), 
while for $d\!=\!4$ this gives an equation that characterizes 
the generating functional $\Gamma[g(x)]$: 
\begin{eqnarray}
&&2\Big[\{ \Sloc,\,\Gamma\}\Big]_4 
+4\Big[\{ \Sloc,\,\Sloc,\,\Sloc,\,\Gamma\}\Big]_4\nn
&&=-\frac{1}{2\kappa_5^2}\left(
 \Big[\{ \Sloc,\,\Sloc\}\Big]_4
 +\Big[\{ \Sloc,\,\Sloc,\,\Sloc,\,\Sloc\}\Big]_4 
\right.\nn
&&\hspace{35mm}\hspace{-15mm}+\,\gamma_3R^2+\gamma_4R_{ij}^2
+\gamma_5R_{ijkl}^2\Bigr). 
\end{eqnarray}
{}From this, we can evaluate the trace of the stress tensor 
for the boundary field theory: 
\ba
 \langle T^i_i \rangle_g\equiv 
  {-2\over\sqrt{g}}\,g_{ij}{\delta\Gamma\over\delta g_{ij}}.
\ea
In fact, using the values in (\ref{wphi}), 
we can show that the trace is given by 
\begin{align}
 \langle T^i_i\rangle_g=
  {2l^3\over 2\kappa_5^2}\Biggl[&
  \left({-1\over 24}+{5a\over 3l^2}+{b\over 3l^2}
  +{c\over 3l^2}\right)R^2  +\left( {1 \over 8}-{5a\over l^2}
  -{b\over l^2}-{3c\over 2l^2}\right)R_{ij}^2  \nn 
  &+{c\over 2l^2}R_{ijkl}^2 
  \Biggr]\,.
 \label{anomaly;bulk}
\end{align}
This correctly reproduces the result\footnote
{The authors of Refs.\ \cite{NO} and \cite{BGN} parametrized 
the cosmological constant $\Lambda$ as 
\ba
 \Lambda=-{d(d-1)\over 2 L^2}\,, \n
\ea
so that their $L$ is related to our $l$, the radius of asymptotic AdS, as 
\ba
 l^2=L^2\left[ 
  1-{(d-3) \over (d-1)L^2}\big( d(d+1)a+db+2c\big)\right].\n
\ea
}
obtained in Refs.\ \cite{NO} and \cite{BGN}, 
where the Weyl anomaly was calculated by perturbatively solving 
the equation of motion near the boundary and by looking at 
the logarithmically divergent term, as in Ref.\ \cite{HS;weyl}.

{}For the case of ${\cal N}\!=\!2$ superconformal $USp(N)$ gauge theory 
in four dimensions, 
we choose $2\kappa_5^2$ such that 
\begin{equation}
 {1\over 2\kappa_5^2}=
  {{\rm Vol}(S^5/\bZ_2)\,({\rm radius~of~}S^5/\bZ_2)^5\over 2\kappa^2},
 \label{5newton}
\end{equation}
where $2\kappa^2=(2\pi)^7g_s^2$ is the ten-dimensional Newton constant 
\cite{Pol}, 
and the radius of $S^5/\bZ_2$ could be set to 
$(8\pi g_sN)^{1/4}$ \cite{APTY}. 
In this relation, we note the replacement $N \to 2N$ as compared to
the $AdS_5\times S^5$ case. 
This is because here we must quantize 
the RR $5$-form flux over $S_5/\bZ_2$ instead of over $S^5$ \cite{N=2CFT}. 
For the $AdS_5$ radius $l$, 
we may also set $l=(8\pi g_s N)^{1/4}$. 
Setting the values $a=b=0$ and 
$c/2l^2=1/32N+{\cal O}(1/N^2)$, as determined in Ref.\ \cite{BGN}, 
we find that the Weyl anomaly (\ref{anomaly;bulk}) takes the form 
\ba
 \langle T^i_i\rangle_g&\!\!\!=\!\!\!&
  {N^2\over 2\pi^2}\left[ 
  \left({-1\over 24}+{1\over 48N}\right)R^2
  +\left({1\over 8}-{3\over 32N}\right)R_{ij}^2
  +{1\over 32N}R_{ijkl}^2\right] 
 +{\cal O}(N^0)\,.
\ea
This is different from the field theoretical result \cite{Duff;Weyl}, 
\ba
 \langle T^i_i\rangle_g&\!\!\!=\!\!\!&
  {N^2\over 2\pi^2}\left[ 
  \left({-1\over 24}-{1\over 32N}\right)R^2
  +\left({1\over 8}+{1\over 16N}\right)R_{ij}^2
  +{1\over 32N}R_{ijkl}^2\right] 
 +{\cal O}(N^0)\,.
\label{anomaly;bgn}
\ea
As was pointed out in Ref.\ \cite{BGN}, 
the discrepancy could be accounted for 
by possible corrections to the radius $l$ 
as well as to the five-dimensional Newton constant. 
In fact, if these corrections are 
\ba
 l=(8\pi g_s N)^{1/4}\left(1+\frac{\xi}{N}\right)\,,\quad
 \frac{1}{2\kappa_5^2}= 
  \frac{{\rm Vol}(S^5/\bZ_2)\,(8\pi g_s N)^{5/4}}{2\kappa^2}
  \left( 1+{\eta\over N}\right)\,,
\ea
then the field theoretical result is correctly reproduced 
for $3\xi+\eta=5/4$.

%
\resection{Conclusion}

In this paper, we investigated higher-derivative gravity systems 
in the context of the AdS/CFT correspondence. 
Although higher-derivative gravity requires more boundary conditions 
than Einstein gravity, 
we pointed out that by choosing the Neumann boundary conditions 
for higher-derivative modes, 
the classical action can be made such that it depends 
only on the boundary values of bulk fields. 
We further derived a Hamilton-Jacobi-like equation 
that determines such a classical action. 
Using this equation, we computed the $1/N$ correction to the 
Weyl anomaly of ${\cal N}\!=\!2$ $G\!=\!USp(N)$ superconformal 
field theory in four dimensions 
on the basis of the holographic description in terms of type IIB string 
theory on $AdS_5\times S^5/\bZ_2$ \cite{N=2CFT}. 
We found that the resulting Weyl anomaly correctly reproduces 
the holographic Weyl anomaly given in Refs.\ \cite{NO} and \cite{BGN}, 
and is consistent with the field theoretical result 
if we take into account the possible corrections 
discussed in Ref.\ \cite{BGN}.

Finally, we comment on how our Neumann boundary condition 
$P\!=\!0$ can be interpreted in the context of the holographic RG. 
To this end, we consider a toy model with the Lagrangian 
\ba
 L={1\over 2}\dot{q}^2+{1\over 2}\mu^2q^2+{c\over 2}\ddot{q}^2\,,
\ea
whose first-order form reads 
\ba
 L=p\dot{q}+P\dot{Q}-H(q,Q;\,p,P), 
\ea
with
\ba
 H(q,Q;\,p,P)=-{1\over 2}\mu^2q^2-{1\over 2}Q^2+Qp+{1\over 2c}P^2. 
\ea
By performing an almost diagonal canonical transformation, 
\ba
 \left( 
  \begin{array}{c}
  q \\
  Q \\
  p \\
  P \\
  \end{array}
 \right) =
 \left(
  \begin{array}{cccc}
  a_1 & a_2 & {1\over m^2}a_3 & {1\over M^2}a_4 \\
  a_3 & a_4 & a_1 & a_2                       \\
  c\,M^2 a_3 & c\,m^2 a_4 & c\,M^2 a_1 & c\,m^2 a_2 \\
  c\,m^2 a_1 & c\,M^2 a_2 & c\,a_3 & c\,a_4 
  \end{array}
 \right)
 \left( 
  \begin{array}{c}
  q^{\prime} \\
  Q^{\prime} \\
  p^{\prime} \\
  P^{\prime} \\
  \end{array}
 \right),
\ea
with 
\ba
 m^2&\!\!\!=\!\!\!&{1-\sqrt{1-4c\,\mu^2}\over 2c}=\mu^2(1+{\cal O}(c))\,,\nn
 M^2&\!\!\!=\!\!\!&{1+\sqrt{1-4c\,\mu^2}\over 2c}={1\over c}(1+{\cal O}(c)),
 \label{toy;mass}
\ea
\ba
 a_1^2={1\over m^2}a_3^2+{1\over 1-2c\mu^2}\,,\qquad
 a_2^2={1\over M^2}a_4^2-{1\over 1-2c\mu^2}\,, 
\ea
the Lagrangian can be rewritten into the following form 
with normalized kinetic term: 
\ba
 L=p^{\prime}\dot{q}^{\prime}+P^{\prime}\dot{Q}^{\prime}
  -H^{\prime}(q^{\prime},p^{\prime};\,Q^{\prime},P^{\prime}), 
\ea
where 
\ba
 H^{\prime}(q^{\prime},Q^{\prime};\,p^{\prime},P^{\prime})=
  {1\over 2}p^{\prime 2}+{1\over 2}P^{\prime 2}
  -{1\over 2}m^2q^{\prime 2}-{1 \over 2}M^2Q^{\prime 2}.
\ea
Since a bulk mode with mass $M$ is coupled to a scaling operator 
with scaling dimension $\Delta={1\over 2}\left(d+\sqrt{d+4M^2}\right)$ 
\cite{GKP}\cite{W;holography}, 
the relation (\ref{toy;mass}) shows that 
the mode $Q^{\prime}$ is coupled to a highly irrelevant operator 
with large scaling dimension when $c\!\ll\! 1$. 
The essential point of this conclusion does not change 
even if the variable $q$ corresponds to 
a bulk field with spin.

Turning to higher-derivative gravity systems, 
the above example shows that $K_{ij}~(\!\sim\! Q\!\sim\! Q')$ 
is highly irrelevant in the dual CFT 
and is approximated well by assuming that it takes a constant value 
along the renormalized trajectory, 
as long as we consider the vicinity of the conformal fixed point. 
This is equivalent to demanding that the corresponding beta function 
vanishes along the renormalized trajectory. 
Since $P^{ij}$, the conjugate momentum of $K_{ij}$, 
can be regarded as the RG beta function of $K_{ij}$, 
this leads to our requirement, $P^{ij}\!=\!0$. 
The holographic RG structure in higher-derivative systems 
will be explored in more detail in a subsequent paper \cite{FMS;fu}.

\section*{Acknowledgments}

The authors would like to thank M.\ Ninomiya, S.\ Nojiri and S.\ Ogushi 
for useful discussions.
The work of M.F. is supported in part by a Grand-in-Aid
for Scientific Research from the Ministry of Education, Science,
Sports and Culture,
and the work of T.S.\ is supported in part
by JSPS Research Fellowships for Young Scientists.

\appendix

\resection{Proof of Theorem}

In this appendix, we give a detailed proof of 
Theorem, \eq{theorem1} and \eq{theorem2}, 
for the action  
\ba
 \bS=\int^t_{t'} dr \Big[ L_0(q^i,\dot{q}^i)
  +c\,L_1(q^i,\dot{q}^i,\ddot{q}^i)\Big], 
\ea
where $i$ runs over some values.  
In the following discussion, 
we focus only on the upper boundary, for simplicity.

We first rewrite the zero-th order Lagrangian $L_0$ 
into the first-order form 
by introducing the conjugate momentum $p_{0i}$ of $q^i$ as  
\ba
 \bS[q(r),p_0(r)]
  =\int^t dr \Big[ p_{0i}\dot{q}^i-H_0(q,p_0)+c\,L_1(q,\dot{q},\ddot{q})
\Big], 
\ea
through the Legendre transformation from $(q,\dot{q})$ to 
$(q,p_0)$ defined by 
\ba
 p_{0i}={\partial L_0\over\partial\dot{q}^i}(q,\dot{q})\,. 
\ea
{}From this, the equation of motion for $p_{0i}$ and $q^i$ is given
by 
\ba
\dot{q}^i&\!\!\!=\!\!\!&{\partial H_0\over\partial p_{0i}}, 
\label{h1;ap} \\
\dot{p_{0i}}&\!\!\!=\!\!\!&-{\partial H_0\over\partial q^i}
+c\left[{\partial L_1\over\partial q^i}
-{d\over dr}\left({\partial L_1\over\partial\dot{q}^i}\right)
+{d^2\over dr^2}\left({\partial L_1\over\partial\ddot{q}^i}\right)
\right]. 
\label{h2;ap}
\ea
Let $\bar{q}(r),\,\bar{p}_0(r)$  be the solution to this equation of motion 
that satisfies the boundary condition 
\ba
 \bar{q}^i(r\!=\!t)=q^i\,. 
\ea
Since this condition determines the classical trajectory uniquely 
[together with the lower boundary values $\bar{q}^i(r\!=\!t')=q^{\prime\,i}
$ 
that we have not written here explicitly], 
the boundary value of $\bar{p}_{0}$ is completely specified by $t$ and $q$: 
$\bar{p}_0(r\!=\!t)\!=\!p_0(t,q)$. 
By plugging the classical solution into the action $\bS$, the 
classical action is obtained as a function of the boundary value $q^i$ 
and $t$: 
\ba
 S(t,q)=\bS[\bar{q}(r),\bar{p}_0(r)].
\ea
In order to derive a differential equation that determines $S(t,q)$, 
we then take the variation of $S(t,q)$. 
Using (\ref{h1;ap}) and (\ref{h2;ap}), this is easily evaluated to be 
\ba
\delta S&\!\!\!=\!\!\!&
\delta t\Big[p_{0i}\dot{q}^i-H_0(q,p_0)+c\,L_1(q,\dot{q},\ddot{q})\Big] \nn
&\!\!\!\!\!\!&+\,\delta\bar{q}^i(t)\left[ p_{0i}+c\left( 
{\partial L_1 \over\partial\dot{q}^i}(q,\dot{q},\ddot{q})-
{d\over dr}\!\left.\left( {\partial L_1 \over\partial\ddot{q}^i}
 (\bar{q},\dot{\bar{q}},\ddot{\bar{q}})\right)\right|_{r=t}
  \right)\right] \nn
&\!\!\!\!\!\!&
 +\,c\,\delta\dot{\bar{q}}^i(t)\,{\partial L_1\over\partial\ddot{q}^i}
  (q,\dot{q},\ddot{q}), 
\label{var}
\ea
where 
\ba
 \dot{q}^i\equiv{d\bar{q}^i\over dr}(r\!=\!t),\qquad 
 \ddot{q}^i\equiv{d^2\bar{q}^i\over dr^2}(r\!=\!t)\,,
\ea
and $\delta\bar{q}^i(t)$ and $\delta\dot{\bar{q}}^i(t)$ are understood 
to be $\delta\bar{q}^i(r)|_{r=t}$ 
and $d\,\delta\bar{q}^i(r)/d r|_{r=t}$, respectively. 
By expanding the classical solution $\bar{q}^i(r)$ around $r\!=\!t$, 
we find that the variations $\delta \bar{q}^i(t)$ 
and $\delta \dot{\bar{q}}^i(t)$ are given by 
\ba
 \delta\bar{q}^i(t)=\delta q^i-\dot{q}^i\,\delta t, \qquad
  \delta\dot{\bar{q}}^i(t)=\delta \dot{q}^i-\ddot{q}^i\,\delta t. 
\ea
Here it is important to note that 
$\dot{q}$ can be written in terms of $q$ and $t$, since 
the classical solution is determined uniquely by the boundary value
$q$. 
Actually it can be shown that 
\ba
\delta\dot{q}^i&\!\!\!=\!\!\!&
{\partial^2H_0\over\partial q^j\partial p_{0i}}\,\delta q^j
+{\partial^2 H_0\over\partial p_{0i}p_{0j}}\,\delta p_{0j} \nn
&\!\!\!=\!\!\!&
{\partial^2H_0\over\partial q^j\partial p_{0i}}\,\delta q^j
+{\partial^2 H_0\over\partial p_{0i}p_{0j}}\,\left( 
{\partial p_{0j}\over\partial t}\delta t
+{\partial p_{0j}\over\partial q^k}\delta q^k\right), 
\ea
where we have used (\ref{h1;ap}) as well as the fact that 
$p_{0}=p_{0}(t,q)$. 
From these relations, 
the variation (\ref{var}) is found to be 
\ba
\delta S=p_i\,\delta q^i-\widetilde{H}(q,p)\,\delta t, 
\ea
with
\ba
 p_i&\!\!\!=\!\!\!&p_{0i} 
 +c\left[ {\partial L_1\over\partial\dot{q}^i}(q,\dot{q},\ddot{q})
  -{d\over dr}\!
  \left.\left( {\partial L_1\over\partial\ddot{q}^i}
  (\bar{q},\dot{\bar{q}},\ddot{\bar{q}})\right)\right|_{r=t}\right.\nn
 &\!\!\!\!\!\!&\left.
  +\,{\partial L_1\over\partial\ddot{q}^j}\left(
  {\partial^2H_0\over\partial q^i\partial p_{0j}}
  +{\partial^2H_0\over\partial p_{0j}\partial p_{0k}}
  {\partial p_{0k}\over\partial q^i}\right)\right], 
 \label{p;ap}\\
 \tH(q,p)&\!\!\!=\!\!\!&H_0(q,p_0) \nn
  &&+\,c\left[ 
  -L_1(q,\dot{q},\ddot{q})
  +\dot{q}^i\left(
  {\partial L_1\over\partial\dot{q}^i}(q,\dot{q},\ddot{q})
  -{d\over dr}\!\left.\left(
   {\partial L_1\over\partial\ddot{q}^i}(\bar{q},\dot{\bar{q}},\ddot{\bar
  {q}})
   \right)\right|_{r=t}\right)
  \right. \nn
&&~~~~~\,\left. +\,
{\partial L_1\over\partial\ddot{q}^i}\left(
\ddot{q}^i
-{\partial^2H_0\over\partial p_{0i}\partial p_{0j}}
{\partial p_{0j}\over\partial t}
\right)\right].
\ea
In order to compute $\tH(q,p)$, we first note that the Hamilton
equation appearing in (\ref{h1;ap}) and (\ref{h2;ap}) gives the relation 
\ba
\ddot{q}^i=
{\partial^2H_0\over\partial p_{0i}\partial q^j}
{\partial H_0\over\partial p_{0j}}
+{\partial^2 H_0\over\partial p_{0i}\partial p_{0j}}
\left( {\partial p_{0j}\over\partial q^k}
{\partial H_0\over\partial p_{0k}}
+{\partial p_{0k}\over\partial t}\right)\,.
\ea
It is then easy to verify that $\tH(q,p)$ takes the form 
\ba
\tH(q,p)=H_0(q,p)-c\,L_1(q,\dot{q},\ddot{q})+{\cal O}(c^2). 
\ea
Here $\dot{q}^i$ and $\ddot{q}^i$ in $L_1$ can be replaced by 
\ba
 f_1^i(q,p)\equiv\left\{H_0(q,p),q^i\right\} 
  =\frac{\partial H_0}{\partial p_i}(q,p) 
\ea
and
\ba
 f_2^i(q,p)&\equiv&\left\{H_0(q,p),\left\{H_0(q,p),q^i\right\}\right\}\nn
 &=&{\partial^2 H_0\over\partial p_i\partial q^j}(q,p)
  {\partial H_0\over\partial p_j}(q,p)
  -{\partial^2 H_0\over\partial p_i\partial p_j}(q,p)
  {\partial H_0\over\partial q^j}(q,p)\,,
\ea
respectively, up to ${\cal O}(c^2)$. 
This completes the proof of (\ref{theorem1}) and (\ref{theorem2}).

%
\section{ADM Decomposition}
\setcounter{equation}{0}

In this appendix, we summarize the components of the Riemann tensor, 
Ricci tensor and scalar curvature written in terms of the ADM 
decomposition. 

In the ADM decomposition, the metric takes the form 
\ba
 ds^2&=&\widehat{g}_{\mu\nu}\,dX^\mu dX^\nu \nn
 &=&N(x,r)^2 dr^2+g_{ij}(x,r)\Bigl(dx^{i}+\lambda^i(x,r)dr\Bigr) 
 \Bigl(dx^{j}+\lambda^{j}(x,r)dr\Bigr).
\ea
Here we use the following basis instead of the 
coordinate basis $\partial_{\mu}$: 
\ba
\widehat{e}_{\widehat{n}}={1\over N}(\partial_r-\lambda^i\partial_i,),\qquad
\widehat{e}_i=\partial_i. 
\ea
In this basis, the components of the metric are given by 
\ba
\left( 
\begin{array}{cc}
 \widehat{g}(\widehat{e}_{\widehat{n}},\widehat{e}_{\widehat{n}}) & 
 \widehat{g}(\widehat{e}_{\widehat{n}},\widehat{e}_j) \\
 \widehat{g}(\widehat{e}_j,\widehat{e}_{\widehat{n}}) & 
 \widehat{g}(\widehat{e}_i,\widehat{e}_j) 
 \end{array}
 \right)
=
\left( 
\begin{array}{cc}
1 & 0 \\
0 & g_{ij}
\end{array}
\right).
\ea
{}For the purpose of computing the Riemann tensor in this basis, it is 
useful to start with the formula 
\ba
\hR^{\sigma}_{\,\,\,\rho\mu\nu}\,\widehat{e}_{\sigma}&\!\!\!=\!\!\!&
\hR(\widehat{e}_{\mu},\widehat{e}_{\nu})\widehat{e}_{\rho} \nn
&\!\!\!=\!\!\!&\left[ \hnab_{\widehat{e}_\mu},
\hnab_{\widehat{e}_\nu} \right]\widehat{e}_{\rho}
-\hnab_{[\widehat{e}_{\mu},\widehat{e}_{\nu} ]}\,\widehat{e}_{\rho}. 
\ea
Each component can be calculated explicitly 
by using the equations 
\ba
\hnab_{\widehat{e}_i}\widehat{e}_j&\!\!\!=\!\!\!&
-K_{ij}\widehat{e}_{\widehat{n}}+\Gamma^k_{ij}\,\widehat{e}_k, \nn
\hnab_{\widehat{e}_i}\widehat{e}_{\widehat{n}}&\!\!\!=\!\!\!&
K_i^k\,\widehat{e}_k, \nn
\hnab_{\widehat{e}_{\widehat{n}}}
\widehat{e}_j&\!\!\!=\!\!\!&
{1\over N}\,\partial_jN\,\widehat{e}_{\widehat{n}}+
\left(K^k_j+{1\over N}\,\partial_j\lambda^k\right)\widehat{e}_k, \nn
\hnab_{\widehat{e}_{\widehat{n}}}
\widehat{e}_{\widehat{n}}&\!\!\!=\!\!\!&
-{1\over N}\,g^{kl}\,\partial_k N\,\widehat{e}_l, \nn
\left[ \widehat{e}_{\widehat{n}}, \widehat{e}_{i}\right]&\!\!\!=\!\!\!&
{1\over N}\,\partial_iN\,\widehat{e}_{\widehat{n}}+
{1\over N}\,\partial_i\lambda^k\,\widehat{e}_{k}, 
\ea
where $K_{ij}$ is the extrinsic curvature and $\Gamma^i_{jk}$ is 
the affine connection with respect to $g_{ij}$. 
We thus obtain 
\ba
\hR_{ijkl}&\!\!\!=\!\!\!&R_{ijkl}-K_{ik}K_{jl}+K_{il}K_{jk}, \nn
\hR_{\widehat{n}jkl}&\!\!\!=\!\!\!&\nabla_lK_{jk}-\nabla_kK_{jl}, \nn
\hR_{\widehat{n}j\widehat{n}l}&\!\!\!=\!\!\!&(K^2)_{jl}-L_{jl},
\ea
with
\ba
L_{ij}={1\over N}\left( 
\dot{K}_{ij}-\lambda^k\,\nabla_k K_{ij}-\nabla_i\lambda^k\,K_{kj}
-\nabla_j\lambda^k\,K_{kj}+\nabla_i\nabla_j N\right). 
\ea
The components of the Ricci tensor 
$\widehat{R}_{\mu\nu}\equiv \widehat{R}^{\rho}_{\,\,\,\mu\rho\nu}
=\widehat{R}_{\nu\mu}$ are given by 
\ba
\hR_{ij}&\!\!\!=\!\!\!&R_{ij}+2(K^2)_{ij}-KK_{ij}-L_{ij}, \nn
\hR_{i\widehat{n}}&\!\!\!=\!\!\!&\nabla^kK_{ki}-\nabla_iK, \nn
\hR_{\widehat{n}\widehat{n}}&\!\!\!=\!\!\!&K_{ij}^2-g^{ij}L_{ij}\,,
\ea
and the scalar curvature is 
\ba
\hR=R+3K_{ij}^2-K^2-2g^{ij}L_{ij}.
\label{scalar;ap}
\ea

%
\section{Boundary Terms}
\setcounter{equation}{0}

In this appendix, we supplement the discussion of
the possible boundary terms given in \S3.

We first consider the infinitesimal transformation 
\ba
 x^i\rightarrow x^{\prime i}=x^{i}+\epsilon^i(x,r),\qquad 
  r\rightarrow r^{\prime}=r+\epsilon(x,r). 
 \label{diffeo;inf}
\ea
Under this transformation, $N,\lambda_i$ and $g_{ij}$ are found 
to transform as 
\ba
{1\over N^{\prime}}&\!\!\!=\!\!\!&{1\over N}(1+\dot{\epsilon}
-\lambda^i\partial_i\epsilon), \nn
\lambda^{\prime}_i&\!\!\!=\!\!\!&\lambda_i-\partial_i\epsilon^j\lambda_j
-\dot{\epsilon}\lambda_i-\partial_i\epsilon\,(N^2+\lambda^2)
-g_{ij}\dot{\epsilon}^j, \nn
g^{\prime}_{ij}&\!\!\!=\!\!\!&g_{ij}-\partial_i\epsilon^kg_{kj}
-\partial_j\epsilon^kg_{ik}-\partial_i\epsilon\,\lambda_j
-\partial_j\epsilon\,\lambda_i. 
\ea
{}Furthermore, $\Gamma^i_{jk}$, the affine connection defined by $g_{ij}$, 
transforms under the diffeomorphism (\ref{diffeo;inf}) as 
\ba
\Gamma^{\prime i}_{jk}=\Gamma^i_{jk}-\partial_j\,\partial_k\epsilon^i
+\Gamma^m_{jk}\,\partial_m\epsilon^i
-\Gamma^i_{mk}\,\partial_j\epsilon^m
-\Gamma^i_{jm}\partial_k\epsilon^m+\tilde{\delta}\Gamma^i_{jk}, 
\ea
with 
\ba
\tilde{\delta}\Gamma^i_{jk}=-\lambda^i\nabla_j\nabla_k\epsilon 
-\partial_j\epsilon\nabla_k\lambda^i-\partial_k\epsilon\nabla_j\lambda^i
-Ng^{il}(\partial_j\epsilon\,K_{lk}+\partial_k\epsilon\,K_{lj}
-\partial_l\epsilon\,K_{jk}). 
\ea
Note that $\tilde{\delta}\Gamma^i_{jk}$ does not contain $\epsilon^i$. 
{}From these relations, it is straightforward to verify that 
the extrinsic curvature transforms as 
\begin{align}
K^{\prime}_{ij}=&\,K_{ij}
-\partial_i\epsilon^l\,K_{lj}-\partial_k\epsilon^l\,K_{jl} \nn
&+N\nabla_i\nabla_j\epsilon
+\partial_i\epsilon\,(\partial_jN-\lambda^lK_{jl})
+\partial_j\epsilon\,(\partial_iN-\lambda^lK_{lj}). 
\end{align}
We can also show that the Riemann curvature $R^i_{\,\,jkl}$ 
transforms under (\ref{diffeo;inf}) as 
\ba
R^{\prime i}_{\,\,\,jkl}&\!\!\!=\!\!\!&R^i_{\,\,jkl}
+\partial_m\epsilon^i\,R^m_{\,\,jkl}
-\partial_j\epsilon^m\,R^i_{\,\,mkl}
-\partial_k\epsilon^m\,R^i_{\,\,jml}
-\partial_l\epsilon^m\,R^i_{\,\,jkm} \nn
&&-\partial_k\epsilon\,\dot{\Gamma}^i_{lj}
+\partial_l\epsilon\,\dot{\Gamma}^i_{kj}
+\nabla_k\tilde{\delta}\Gamma^i_{lj}
-\nabla_l\tilde{\delta}\Gamma^i_{kj}. 
\ea

As argued in \S $3$, 
we focus on the diffeomorphism that obeys the condition 
(\ref{diffeo;boundary}). 
This is equivalent to the following relation in an infinitesimal form: 
\ba
 \partial_i\epsilon(r\!=\!r_0)=0. 
 \label{r0;ap}
\ea
Therefore, we find that the boundary action (\ref{Sb}) is invariant under
this diffeomorphism.

We remark that in the above, we have discarded boundary terms of the form 
\ba
 \bS'_b=\int_{\Sigma_d}d^d x \sqrt{g}\left(K^{ij}L_{ij}
  +K g^{ij}L_{ij}\right),
\ea
although these are allowed by the diffeomorphism.\footnote
{By definition, the $(d+1)$-dimensional scalar curvature 
$\widehat{R}$ is a scalar. 
It thus follows from (\ref{scalar;ap}) that $L_{ij}(r\!=\!r_0)$ 
transforms as a tensor under the diffeomorphism with (\ref{r0;ap}). 
}
The reason is that if there were such boundary terms, 
they would require us to further introduce an extra boundary condition, 
since  
\ba
 \delta \bS'_b = \int_{\Sigma_d}d^d x \sqrt{g}
  \left[\cdots +\delta\dot{K}_{ij}P_2^{ij}(g_{kl},K_{kl})\right].
\ea

%
\section{Coefficients in Eq.\ (\ref{ham})}
\setcounter{equation}{0}

We have the following values for the coefficients in Eq.\ (\ref{ham}):
\ba
A_1 &\!\!\!=\!\!\!& \frac{2b-x_2}{2(b+4c)}, \quad 
A_2 = {\frac {4b+8c-3{x_5}}{2(b+4c)}}, \quad 
A_3 = -{\frac {b+{x_4}}{b+4c}}, \nn
A_4 &\!\!\!=\!\!\!& -{\frac {4ab-16ac+b{x_1}
+4c{x_1}-4a{x_2}+2{b}^{2}-b{x_2}}{2(b+4c)(4da+(d+1)b+4c)}}, \nn
A_5 &\!\!\!=\!\!\!& -\frac{4ab-16ac+2{b}^{2}+b{x_4}+4c{x_4}-12a{x_5}-3b{x_
5}}
{2(b+4c)(4da+(d+1)b+4c)}, \nn
A_6 &\!\!\!=\!\!\!&\frac{4ab-16ac-3b{x_3}-12c{x_3}+8a{x_4}+2{b}^{2}+2b{x_4}}
{2(b+4c)(4da+(d+1)b+4c)}, 
\ea
\ba
B_1 &\!\!\!=\!\!\!& \frac{1}{4(b+4c)(4da+(d+1)b+4c)} \nn
&\!\!\!\!\!\!&\times\Big[4{b}^{3}+4(d+1)a{b}^{2}+4a{x_2}^{2}-4{b}^{2}{x_2}
+b{x_2}^{2}+64a{c}^{2}-8ab{x_2} \nn
&\!\!\!\!\!\!&+16(d-2)abc-4dc{{x_1}}^{2}
-db{x_1}^{2}+4{b}^{2}{x_1}+16bc{x_1}-8c{x_1}{x_2} \nn
&\!\!\!\!\!\!&+32ac{x_2}-2b{x_1}{x_2}
+8dab{x_1}+32dac{x_1}\Big], \nn
B_2 &\!\!\!=\!\!\!& \frac{16bc+4b{x_2}-{x_2}^{2}}{4(b+4c)}, \quad
B_3 = c, 
\ea
\ba
C_1 &\!\!\!=\!\!\!& \frac{1}{4(b+4c)(4da+(d+1)b+4c)} \nn
&\!\!\!\!\!\!&\times\Big[8{b}^{3}-8ab{x_2}-16(d+1)bcx_1-64c^2x_1-32dacx_1 
\nn
&\!\!\!\!\!\!& -4db^2x_1+8dab{x_1}-4{b}^{2}{x_2}
+32ac{x_2}+8\,dab{x_4}-24ab{x_5} \nn 
&\!\!\!\!\!\!& +24a{x_2}{x_5}+6b{x_2}{x_5}-12{b}^{2}{x_5}
+32(d-2)abc-2db{x_1}{x_4} \nn
&\!\!\!\!\!\!&+8(d+1)ab^2+16bc{x_4}+4{b}^{2}{x_4}
-2b{x_2}{x_4}-6b{x_1}{x_5}\nn 
&\!\!\!\!\!\!&-8c{x_2}{x_4}-8dc{x_1}{x_4}+32dac{x_4}-24c{x_1}{x_5}
+96ac{x_5}+128a{c}^{2}\Big], \\
C_2 &\!\!\!=\!\!\!& \frac{1}{4(b+4c)(4da+(d+1)b+4c)} \nn
&\!\!\!\!\!\!&\times\Big[-16{b}^{2}c+8bcx_2+64c^2x_1+32dacx_1
-32(d+2)abc \nn
&\!\!\!\!\!\!&-8(7d+5)ab^2-(d-3)b{x_2}{x_4}-4(d-4)a{x_2}{x_4} 
\nonumber
\ea
\ba
&\!\!\!\!\!\!&+8(d-2)ab{x_4}+2(d-3){b}^{2}{x_4}
+3(d-1)b{x_2}{x_3}-6(d-1){b}^{2}{x_3}\nn
&\!\!\!\!\!\!&-4(d+3){b}^{3}+32acx_2+24(d+1)abx_2+16(d+1)bcx_1 \nn
&\!\!\!\!\!\!&+64d{a}^{2}{x_2}-12c{x_2}{x_3}+2(d+3)b^2x_2+8dabx_1+4db^2x_1
\nn
&\!\!\!\!\!\!&-6db{x_1}{x_3}+24bc{x_3}-4b{x_1}{x_4}+12da{x_2}{x_3}+8bc{x_4}+
96dac{x_
3}\nn
&\!\!\!\!\!\!&-4c{x_2}{x_4}-16c{x_1}{x_4}+64ac{x_4}-24dc{x_1}{x_3}-128d{a}^
{2}b
-128a{c}^{2}\Big], \nn
C_3 &\!\!\!=\!\!\!&
\frac{32bc+6b{x_5}-3{x_2}{x_5}+64{c}^{2}-8c{x_2}}
{2(b+4c)}, \nn
C_4 &\!\!\!=\!\!\!& \frac{-8bc+2b{x_4}-2b{x_1}-{x_2}{x_4}-8c{x_1}
+4c{x_2}}{b+4c}, \nn
C_5 &\!\!\!=\!\!\!& -12c-2{x_2}, 
\ea
\ba
D_1 &\!\!\!=\!\!\!& \frac{{8bc-9{x_5}^{2}-48c{x_5}-32{c}^{2}}}{4(b+4c)}, 
\nn
D_2  &\!\!\!=\!\!\!&
\frac{-4b{x_5}-16bc-16c{x_4}-6{x_4}{x_5}+8c{x_5}}
{2(b+4c)},
\nn
D_3 &\!\!\!=\!\!\!& \frac{1}{4(b+4c)(4da+(d+1)b+4c)} \nn
&\!\!\!\!\!\!&\times\Big[
-6b{x_4}{x_5}-64{c}^{2}{x_4}+96ac{x_5}
-16(d+1)bc{x_4}-32dac{x_4}+128{c}^{3}\nn
&\!\!\!\!\!\!&-4d{b}^{2}{x_4}-24c{x_4}{x_5}
+32(d+2)b{c}^{2}-db{x_4}^{2}
-4dc{x_4}^{2}+8(d+1){b}^{2}c \nn
&\!\!\!\!\!\!&+4{b}^{3}+64(2d+1)a{c}^{2}
+4(d+1)a{b}^{2}+16(3d-2)abc-24ab{x_5} \nn
&\!\!\!\!\!\!&-12{b}^{2}{x_5}-8dab{x_4}+9b{x_5}^{2}+36a{{x_5}}^{2}\Big],
\nn
D_4 &\!\!\!=\!\!\!& \frac{1}{4(b+4c)(4da+(d+1)b+4c)} \nn
&\!\!\!\!\!\!&
 \times\Big[-8{b}^{3}-32c{{x_4}}^{2}+48a{x_4}{x_5}
-16da{{x_4}}^{2}+24ab{x_5}+12{b}^{2}{x_5}+12b{x_4}{x_5} \nn
&\!\!\!\!\!\!&
-24dab{x_3}-96dac{x_3}+64{c}^{2}{x_4}-96ac{x_5}-192{c}^{2}{x_3} \nn
&\!\!\!\!\!\!&
-72c{x_3}{x_5}+32(d+2)abc-48(d+1)bc{x_3}-8(3d+2)ab{x_4} \nn
&\!\!\!\!\!\!&
+16(d-1)bc{x_4}+32(d+2)ac{x_4}+16(d+1){b}^{2}c-4(d+2)b{{x_4}}^{2} \nn
&\!\!\!\!\!\!&
-4(d+4){b}^{2}{x_4}-8(d+1)a{b}^{2}-128a{c}^{2}-6db{x_3}{x_4} \nn
&\!\!\!\!\!\!&
-18b{x_3}{x_5}-24dc{x_3}{x_4}-12d{b}^{2}{x_3}+64b{c}^{2}\Big], \nonumber
\ea
\ba
D_5 &\!\!\!=\!\!\!& 
\frac{1}{4(b+4c)(4da+(d+1)b+4c)} \nn
&\!\!\!\!\!\!&
\times\Big[
16a{x_4}^{2}+64{c}^{2}{x_3}-8dab{x_3}
-32dac{x_3}-12b{x_3}{x_4}-48c{x_3}{x_4},\nn
&\!\!\!\!\!\!&
+4(d-2){b}^{2}{x_3}-64ac{x_4}+8{b}^{2}{x_4}+4b{x_4}^{2}
+4{b}^{3}+64a{c}^{2}-9db{{x_3}}^{2} \nn
&\!\!\!\!\!\!&-36dc{x_3}^{2}+4(d+1)a{b}^{2}+16(d-2)abc+16ab{x_4}
+16(d-1)bc{x_3}\Big], 
\ea
\ba
E_1 &\!\!\!=\!\!\!& 4b+2{x_1}-{x_2},\nn
E_2  &\!\!\!=\!\!\!& -2b+8c-2{x_2}, \nn
E_3 &\!\!\!=\!\!\!& -8c-{x_2},\nn
E_4 &\!\!\!=\!\!\!& 2b-2{x_1}. 
\ea

%

\end{document}